\documentclass[graybox, envcountchap]{svmult}

\usepackage[square,numbers]{natbib}
\usepackage{mathptmx}        % selects Times Roman as basic font
\usepackage{amsmath}
\usepackage{amssymb}
\usepackage{color}
\usepackage{helvet}          % selects Helvetica as sans-serif font
\usepackage{courier}         % selects Courier as typewriter font
\usepackage{dirtree}
\usepackage{makeidx}        % allows index generation
\usepackage{graphicx}        % standard LaTeX graphics tool
\usepackage{subfig}

\usepackage{multicol}        % used for the two-column index
\usepackage[bottom]{footmisc}% places footnotes at page bottom

\usepackage{hyperref}        %for hyperlinks
\hypersetup{colorlinks=true,urlcolor=blue}

\usepackage[misc]{ifsym}

\def\p{\partial}
\def\msun{M$_{\odot}$}
\def\Msun{M$_{\odot}$ }
\def\be{\begin{equation}}
\def\ee{\end{equation}}
\def\bi{\begin{itemize}}
\def\i{\item}
\def\ei{\end{itemize}}
\def\ben{\begin{enumerate}}
\def\een{\end{enumerate}}
\def\bea{\begin{eqnarray}}
\def\eea{\end{eqnarray}}
\def\bt{\begin{tabbing}}
\def\et{\end{tabbing}}

\def\edo{\end{document}}

\newcommand{\tlg}{\tilde{\gamma}}
\newcommand{\tlG}{\tilde{\Gamma}}
\newcommand{\emfp}{e^{-4\phi}}
\newcommand{\tlA}{\tilde{A}}
\newcommand{\tlR}{\tilde{R}}

\newcommand{\dt}[1]{\partial_t #1}
\newcommand{\pdu}[3]{\partial_{#3} #1^{#2}}
\newcommand{\pdl}[3]{\partial_{#3} #1_{#2}}
\newcommand{\pdpdu}[4]{\partial_{#3}\partial_{#4} #1^{#2}}
\newcommand{\pdpdl}[4]{\partial_{#3}\partial_{#4} #1_{#2}}
\newcommand{\upwindu}[3]{\beta^{#3}\bar{\partial}_{#3} #1^{#2}}
\newcommand{\upwindl}[3]{\beta^{#3}\bar{\partial}_{#3} #1_{#2}}
\newcommand{\tlGn}{\tilde{\Gamma}_{\mathrm{(n)}}}
\newcommand{\tlGmixed}[2]{\tlG_{#1}^{\;\;\;#2}}

\def\rg{$\vec{r}_{\rm G}$}

\def\M4{M$_4$}
\def\M6{M$_6$}
\def\lcm6{LCM$_6$}
\def\qcm6{QCM$_6$}

\def\wh3{W$_{\rm H,3}$}
\def\wh4{W$_{\rm H,4}$}
\def\wh5{W$_{\rm H,5}$}
\def\wh6{W$_{\rm H,6}$}
\def\wh7{W$_{\rm H,7}$}

\newcommand{\SpB}{\texttt{SPHINCS\_BSSN}\; }

\newcommand{\Ma}{\texttt{MAGMA2}\;}

\newcommand{\SPHI}{\texttt{SPHINCS\_BSSN }}
\newcommand{\sphi}{\texttt{SPHINCS\_BSSN}}
\newcommand{\AHFD}{\texttt{AHFinderDirect}}
\newcommand{\QLM}{\texttt{QuasiLocalMeasures}}

\makeindex             % used for the subject index
\begin{document}

%%%%%%%%%%%%%%%%%%%%%%%%%%%%%%%%%%%%%%%%%%%%%%%%%%%%%%%%%%%%%%%%%

\title{SPHINCS\_BSSN: Numerical Relativity with Particles}
\author{Stephan Rosswog and Peter Diener}
\institute{Stephan Rosswog (\Letter) \at Sternwarte Hamburg, Gojenbergsweg 112, 21029 Hamburg, Germany \& The Oskar Klein Centre, Department of Astronomy, Stockholm University, Stockholm, Sweden;  \email{stephan.rosswog@uni-hamburg.de}
\and Peter Diener \at Center for Computation \& Technology, Louisiana State University, 70803, Baton Rouge, LA, USA \& 
Department of Physics \& Astronomy, Louisiana State University, 70803, Baton Rouge, LA, USA \email{diener@cct.lsu.edu}}
\maketitle
\abstract{
In this chapter we describe the {\em Lagrangian} numerical relativity code \sphi.
This code evolves spacetimes in full General Relativity by integrating the BSSN equations
on structured meshes with a simple dynamical mesh refinement strategy.
The fluid is evolved by means of freely moving Lagrangian particles,
that are evolved using a modern Smooth Particle Hydrodynamics (SPH) formulation.
To robustly and accurately capture shocks, our code uses artificial
dissipation terms, but, similar to Finite Volume schemes, we apply a slope-limited
reconstruction within the dissipative terms and we use in addition time-dependent dissipation 
parameters, so that dissipation is only applied where needed. 
The technically most complicated, but absolutely crucial part of the methodology, 
is the coupling between the particles and the mesh. For the mapping of the energy-momentum tensor
$T_{\mu\nu}$ from the particles to the mesh, we use a sophisticated combination of "Local Regression Estimate" (LRE) method and a "multi-dimensional optimal order detection" (MOOD) approach which we describe in some detail.
The mapping of the metric quantities from the 
grid to the particles is achieved by a quintic Hermite interpolation.\\
Apart from giving an introduction to our numerical methods, we
demonstrate the accurate working of our code by presenting a set of representative 
relativistic hydrodynamics tests. We begin with a relativistic shock tube test, 
then compare the frequencies of a fully relativistic neutron star with reference
values from the literature and, finally,  we present full-blown merger simulations 
of irrotational binary systems, one case where a central remnant survives and 
another where a black hole forms, and of a binary where only one of the stars is rapidly spinning.}

%%%%%%%%%%%%%%%%%%%%%%%%%%%%%%%%%%%%%%%%%%%%%%%%%%%%%%%%%

\section{Introduction}
\label{sec:intro}

% advent of GW-astronomy
Despite their early postulation by Albert Einstein in 1916 \cite{einstein16,einstein18},
gravitational waves (GWs) 
remained a topic of sustained debate for more than half a century
\cite{kennefick07}. Their physical reality was only firmly established
after the discovery of the Hulse-Taylor pulsar in 1974 \cite{hulse75}: 
the secular orbital decay of this neutron star binary system is in nearly 
perfect agreement with the prediction from Einstein's theory and this
finally brought the decades' long controversy to an end. Still, the first 
{\em direct} detection of GWs had to wait
until 2015 when both LIGO detectors recorded the last 0.2 seconds
of the inspiral and the subsequent merger of two $\sim 30$ \Msun black holes \cite{abbott16a}.\\
The first detection of a GW source that actually
involved {\em matter} (rather than "just black holes"), followed soon: in August 2017 a $\sim$ minute-long
GW-signal from a merging neutron star binary (GW170817) was detected. Most excitingly, 
this event was also observed in electromagnetic (EM) waves, first as a
short gamma-ray burst (GRB) that followed the peak of the GW emission with a delay of 1.7~s, 
later ($\sim$ days) as a radioactively powered thermal transient peaking 
at optical/near-infrared wavelengths frequently called a "kilonova",
and much later ($\sim$ months) also in radio wavelengths, see \cite{margutti21} for a recent review.\\
This event meant a huge leap forward for many 
long-standing (astro-)physics questions. For example, the electromagnetic 
follow-up observations of GW170817 revealed the actual host galaxy where
the merger took place and this placed the event in an 
astronomical environment and connected it to potential stellar evolution paths 
\cite{levan17}. Since the distance to the host galaxy is known, it was possible to use the 1.7~s delay between GWs and the GRB to
show that GWs propagate to a relative accuracy of $10^{-15}$ at the speed of light \cite{abbott17d}.
 This, in turn, eliminated families of alternative 
theories of gravity where $c_{\rm GW} \ne c$.
GW170817 also demonstrated the exciting possibility to constrain the properties of cold nuclear matter
via the effects of tidal deformability and thereby ruled out some very stiff
equations of state \cite{abbott17c}. The event further
showed that neutron star mergers are, as expected on theoretical 
grounds \cite{lattimer77,symbalisty82,eichler89,rosswog99,freiburghaus99b}, 
major cosmic production sites of heavy "r-process" elements \cite{cowan21}, and that
they are useful probes of the cosmic expansion \cite{abbott17a}. \\
All of these breakthroughs were enabled by the {\em
combined} detection of the same event via different messengers, here
GWs and photons.
This detection splendidly illustrated the huge discovery potential of
multi-messenger astrophysics, but it also underlined the enormous
modelling challenge: in addition to strong-field gravity, relativistic
fluid dynamics, neutrino transport, magnetic fields, nuclear
reactions and atomic physics, also (non-equilibrium) radiative transfer should be included in theoretical models
in order to make the connection to the
EM observations.

% Role of NR-/multi-physics simulations
Numerical simulations, especially numerical relativity simulations \cite{alcubierre08,baumgarte10,rezzolla13a,shibata16},
play a very crucial role in connecting the physical processes during merger
to potentially observable signals. Such simulations have increasingly
become tools for exploring  signatures from both established
and more speculative parts of fundamental physics.
Multi-physics numerical relativity simulations are
  extremely complex and  time consuming: codes typically
  evolve  several dozen variables per spacetime (grid-) point and, depending
  on the simulated physics, many more variables may be needed. For
  example, neutrino transport is needed to make detailed predictions
  for the neutron-to-proton-ratio in the ejecta, which, in turn, determines 
  which elements are formed and therefore also how the electromagnetic signal looks like. But evolving
  several neutrino flavours, with many energy groups for each of them, 
  becomes a serious extra computational burden \cite{mezzacappa20,radice22,foucart23}. To make things worse, the spacetime
  needs to be evolved with very small numerical time steps that are limited
  by the Courant criterion \cite{press92}.
  The "signal speed" that enters this criterion is the speed of light, which limits 
  the admissible time steps to $\Delta_t \approx 10^{-7} \; s
  \; \left( \frac{\Delta x}{100 \rm m}\right)$, where $\Delta x$ is
  the spatial length to be resolved.

To date, essentially all numerical relativity codes that solve the full set of
Einstein equations are {\em Eulerian}, i.e. they solve the fluid equations
on computational (usually structured or adaptive) grids. The only exception 
is our code \SPHI \cite{rosswog21a,diener22a,rosswog22b,rosswog23a}, which is the main topic
of this book chapter. Using an alternative methodology makes it possible to independently
cross-validate the longer established Eulerian methods, but
it also comes with distinctive advantages for multi-messenger astrophysics. 
One of its salient features
is that --contrary to Eulerian methods-- vacuum does not need to be
modelled hydrodynamically: it is simply characterized by the absence of computational
elements (here: particles) and no computational resources need to be spent to evolve it. Moreover, the neutron star surface, which is an
eternal source of numerical trouble for Eulerian methods, is very
well-behaved and does not need any particular treatment compared to
the rest of the fluid\footnote{This is illustrated later, see Fig.~\ref{fig:tov_2M}.}. The ejecta of a neutron star merger correspond
to only $\sim$ 1\% of the binary mass, but they are responsible for
essentially all of the EM emission. Therefore, the accurate prediction
of the properties of this small amount of matter is of paramount
importance for multi-messenger astrophysics. This is, in fact, one of
the key strengths of a code like \sphi: advection is exact (i.e. not
depending on the resolution) and the properties of a particle, say its
electron fraction $Y_e$, are simply transported by the particle unless
they are changed by physical processes (e.g. by an electron capture reaction), 
but no purely numerical "diffusive" processes occur.

In the rest of this chapter, we will describe the methodology behind
our Lagrangian Numerical Relativity code \SPHI
and we will present some of the first applications.  In Sec.~\ref{sec:methods} we will describe how
we evolve fluids and, since in a numerical relativity context the use
of particles is not widespread, we start with some basics of the SPH method.
We further outline how the spacetime is evolved and how the fluid particles and the grid cells for
the spacetime evolution exchange information. In
Sec.~\ref{sec:applications} we will show some applications of
the code including some standard benchmark tests, but also some of the
first astrophysical simulations. We finally will summarize this chapter in
Sec.~\ref{sec:summary}.

\section{Methodology}
\label{sec:methods}
Here we will briefly summarize the main methodological elements that enter in \sphi.
We will focus here on the main ideas and in some cases we will refer to our more technical
papers \cite{rosswog21a,diener22a,rosswog22b,rosswog23a} for the full details.

\subsection{Relativistic Smooth Particle Hydrodynamics}
Our main aim here is to describe how we evolve matter by means of a modern, general relativistic
Smooth Particle Hydrodynamics (SPH) code. But before we do so, we will summarize
some basic SPH techniques and we will briefly discuss the (simpler to understand) Newtonian
SPH approach. Once equipped with this basic knowledge, the general relativistic version
should be straight forward to grasp.\\
This introduction to the topic is by no means meant as a comprehensive SPH review, for a more 
complete introduction to the SPH-method, its variants and subtleties and its various 
applications in astrophysics and cosmology, we refer to extensive reviews that exist in 
the literature \cite{monaghan92,monaghan05,rosswog09b,springel10a,price12a,rosswog15b,rosswog15c}. 
For relativistic formulations of SPH, we specifically want to point to \cite{laguna93a,siegler00a,monaghan01,rosswog09b,rosswog10a,rosswog10b,rosswog11a,rosswog21a,diener22a,rosswog22b,rosswog23a}. Among these, we  will frequently refer to \cite{rosswog09b} 
since here many results are derived very explicitly  in a step-by-step fashion.

\subsubsection{A primer on basic SPH-techniques}
The main idea of SPH is to model a fluid via spherical 
particles of finite size and overlapping support. These particles
move with the local fluid velocity, i.e. SPH is a fully {\em Lagrangian} method,
as opposed to Eulerian or Adaptive Lagrangian Eulerian, or "ALE", methodologies.
The overlapping nature of the particles distinguishes them, for example,
from (usually quasi-) Lagrangian Moving Mesh methods, see e.g. \cite{springel10b,duffell11}, which instead 
tessellate space into non-overlapping cells which have a sharp interface
surface between them. The task is now to translate the continuum equations of 
Lagrangian fluid dynamics into evolution equations for the particles, ideally in a way that ensures that nature's conservation laws are enforced on the level of the particle evolution equations. In the 
simplest case of ideal, non-relativistic fluids the equations read
\bea
\frac{d \rho}{dt}   &=& - \rho \nabla \cdot \vec{v}\label{eq:drhodt}\\
\frac{d \vec{v}}{dt} &=& - \frac{\nabla P}{\rho} + \vec{f}\label{eq:dvdt}\\
\frac{d u}{dt}&=& \frac{dq}{dt} + \frac{P}{\rho^2} \frac{d \rho}{dt}.
\label{eq:dudt}
\eea
Here,
$\rho$ is the mass density, $\vec{v}$ the fluid velocity, $P$ the pressure,
$\vec{f}$ denote other "body forces" e.g. from gravity, $u$ is the specific 
internal energy (i.e. energy per mass) and $dq/dt$ denotes
potential sources of heating and/or cooling. The Lagrangian time derivative, $d/dt$,
is related to the Eulerian time derivative $\p/\p t$ by
\be
\frac{d}{dt} (.)= \left(\frac{\p}{\p t} + \vec{v} \cdot\nabla\right) (.).
\label{eq:Lag_dt}
\ee
Note that Eq.~(\ref{eq:dudt}) is
simply the first law of thermodynamics written "on a per mass basis", i.e. the involved quantities are e.g. energy per mass or volume per mass (=1/density).\\ 
At the heart of the SPH method is kernel interpolation. Assuming that some quantity $f$ is known at particle location 
$b$, a smoothed version of the quantity $\langle f \rangle$ reads
\be
\langle f \rangle (\vec{r})= \int f(\vec{r'}) \; W_h(\vec{r}'-\vec{r})
\; d^3x' \approx \sum_b V_b f_b W_h(\vec{r_b} - \vec{r})= \sum_b 
\frac{m_b}{\rho_b} \; f_b \; W_h(\vec{r_b} - \vec{r}).
\label{eq:kernel_smoothing}
\ee
Here $W_h$ is a smoothing kernel, see below, whose support size is
determined by the length scale $h$, which, in an SPH context, is usually
referred to as the ``smoothing length''. In the approximation we
have applied a one-point quadrature. In SPH, one usually chooses the
particle volume as $m_b/\rho_b$, where $m_b$ and $\rho_b$ are the
particle mass and mass density at the position of particle $b$, 
but other choices are possible, see
e.g. \cite{hopkins13,rosswog15b,cabezon17} for alternatives.
One simple application of Eq.~(\ref{eq:kernel_smoothing}) is to choose the density $\rho$
for the quantity $f$, which yields
\be
\langle\rho\rangle_a= \sum_b m_b W_{ab}(h_a),
\label{eq:rho_sum}
\ee
where $W_{ab}(h_a)$ is a short-hand for $W((\vec{r}_a-\vec{r}_b)/h_a)$.
In other words: the density can be calculated by summing over the
local neighbourhood and weighting each particle's mass according to how
far away it is from particle $a$. If one keeps each particle's mass fixed in time, one has enforced {\em exact} mass conservation and one does not
have to solve the mass conservation equation (\ref{eq:drhodt}) explicitly,
but one can do so if this is desired.\\
Clearly, for the smoothing
procedure in Eq.~(\ref{eq:kernel_smoothing}) to make sense, the smoothing
kernel must have certain properties:
\bi
\i it needs to be normalized, $\int W_h(\vec{r}') d^3r'=1$,
\i obviously, the dimension of the kernel has to be ``1/volume'',
\i it has to have the ``delta-property'', so as to reproduce the 
original function in the limit of vanishing support size, 
\be
\lim_{h\rightarrow 0} \int W_h(\vec{r}'-\vec{r}) \; f(r') \; d^3r' = f(r) 
\ee
\i and it should (at least) have a continuous second derivative.
\ei
A number of other properties are desirable, for example, the kernel
should be "radial", i.e. $W_h(\vec{x}-\vec{x}')=
W_h(|\vec{x}-\vec{x}')|)$, to easily allow for the conservation of
momentum and angular momentum, and its Fourier transform should fall
of rapidly with wave number \citep{monaghan85b}. As a side remark: it may seem
appealing to use an oblate kernel to model, say, an accretion disk, but this
approach would sacrifice one of SPH's most salient features: its exact
angular momentum conservation. The latter can be easily enforced when radial kernel are used, see Sec.~2.4 in \cite{rosswog09b} for an explicit demonstration how radial kernels lead to exact angular momentum conservation.\\
For a gradient estimate, one can straight
forwardly apply the nabla operator to Eq.~(\ref{eq:kernel_smoothing})\footnote{From now
on we drop the distinction between a function and its numerical approximation.}
\be
\nabla f(\vec{r}) = \sum_b \frac{m_b}{\rho_b} \; f_b \; \nabla_r
W_h(\vec{r_b} - \vec{r}),
\label{eq:nabla_f}
\ee
where the index $r$ at the nabla operator indicates that the
derivative is taken with respect to the variable $\vec{r}$ (rather
than $\vec{r}_b$). The derivatives are usually needed at the position
of a particle (here labelled $a$) and abbreviated as
\be
(\nabla f)_a= \sum_b \frac{m_b}{\rho_b} f_b \nabla_a W_{ab}(h_a),
\ee
where  the label at the nabla operator is again a reminder that the derivative
is taken with respect to $\vec{r}_a$ (and not $\vec{r}_b$). From Eq.~(\ref{eq:kernel_smoothing})
it is obvious that one would want to have
the ``partition of unity'' property in a discrete form at any position $\vec{r}$
\bea
1 &=& \sum_b \frac{m_b}{\rho_b} \; W_h(\vec{r_b} - \vec{r})\\
0 &=& \nabla_r (1)= \sum_b \frac{m_b}{\rho_b} \; \nabla_r W_h(\vec{r_b} - \vec{r}),
\eea
where the second equation is simply the result of applying the nabla
operator to the first one. In standard SPH these equations
are only approximately fulfilled and the accuracy depends on the kernel function used,
its support size (or the number of contributing particles) and on the exact particle 
distribution. This has the effect, that when applying Eq.~(\ref{eq:nabla_f})
to  a constant function $f=f_0$, one only approximately recovers the
theoretical value of zero
\be
\nabla_r \left(\sum_b \frac {m_b}{\rho_b} f_b W_h(\vec{r_b} -
  \vec{r})\right)= f_0 \sum_b \frac {m_b}{\rho_b} \nabla_r W_h(\vec{r_b} -
  \vec{r}) \approx 0.
\ee
This, however, can be easily corrected by simply ``subtracting a numerical
zero'', so that a better gradient estimate reads
\begin{align}
\left( \nabla f\right)^{(1)}_a & = \sum_b \frac{m_b}{\rho_b} f_b \nabla_a W_{ab}(h_a)
- \left(f_a \sum_b \frac{m_b}{\rho_b} \nabla_a W_{ab}(h_a)\right) \nonumber \\
 & = \sum_b \frac{m_b}{\rho_b} (f_b-f_a) \nabla_a W_{ab}(h_a),
\label{eq:better_grad}
\end{align}
and this expression vanishes by construction when all $f_k$ are the same. As
an example, we can apply this gradient prescription to find an
estimate for the velocity divergence. This quantity plays an important role in
fluid dynamics, since $\nabla \cdot\vec{v}= 0$ defines ``incompressibility",
while $\nabla \cdot\vec{v}>0$ indicates local expansion and $\nabla
\cdot\vec{v}<0$ means local compression of the fluid. The resulting expression is
\be
(\nabla \cdot \vec{v})_a= \sum_b \frac{m_b}{\rho_b} (\vec{v}_b -
\vec{v}_a) \cdot \nabla_a W_{ab}(h_a).
\ee
An alternative discretization can be found by taking Eq.~(\ref{eq:drhodt})
and calculating the Lagrangian time derivative of Eq.~(\ref{eq:rho_sum}) directly
\be
(\nabla \cdot \vec{v})_a= -\frac{1}{\rho_a} \frac{d\rho_a}{dt}= -\frac{1}{\rho_a} \sum_b m_b \frac{dW_{ab}}{dt}
= \frac{1}{\rho_a} \sum_b m_b (\vec{v}_b - \vec{v}_a) \cdot \nabla_a W_{ab}(h_a),
\label{eq:drho_dt}
\ee
where we have taken the straight-forward Lagrangian time derivative of the kernel function (see the kernel table in Sec.~2.3 in \cite{rosswog09b} where this is shown explicitly). Thus, we have 
derived an alternative numerical discretization for
$\nabla \cdot \vec{v}$, where the difference between the two estimates is whether
the local density ("$\rho_a$") or the weighted nearby densities ("$\rho_b$") are used. 
Both are equally valid and both vanish 
for the uniform velocity case.\\
For easy later reference, we explicitly write down 
once more the  Lagrangian time derivative of the density in Eq.~(\ref{eq:rho_sum}) that we have just found
\be
\frac{d \rho_a}{dt}= \sum_b m_b \frac{d}{dt} \left[W_{ab}(h_a) \right]= \sum_b m_b (\vec{v}_a - \vec{v}_b) \cdot \nabla_a W_{ab}(h_a)
\ee
and we also note that 
\be
\nabla_a \rho_b=  \sum_k m_k \nabla_a W_{bk} (h_b)
\ee
(the detailed steps can be found in Sec.~3.2 of \cite{rosswog09b}). Note 
that we have neglected small terms, usually called "grad-h" terms,
that result from taking derivatives of the kernel with respect to the smoothing lengths.
These terms have been derived for Newtonian hydrodynamics by \cite{springel02,monaghan02}, for special-relativistic SPH in \cite{rosswog10b} and for the General Relativistic case in \cite{rosswog10a}. We had explored the effects of these grad-h terms in \cite{rosswog07c},
but found them to be too close to unity to warrant the extra effort of an additional 
iteration ($\rho$ depends on $h$ and $h$ depends on $\rho$, so one needs an iteration for 
consistent values).  Therefore, we will ignore these "grad-h"-terms for the rest of this 
chapter.

\subsubsection{Newtonian SPH}
We have already seen in Eq.~(\ref{eq:rho_sum}) that the continuity
equation does not need to be solved explicitly, and if we keep the particle masses
fixed, we have enforced exact mass conservation. It now turns out that by insightful choices
(see the discussion in Sec.~2.3 - 2.5 in \cite{rosswog09b}), one can also bring the energy and momentum
equation into discrete forms so that they conserve the total energy and momentum 
exactly. If a radial
kernel is used, then the total angular momentum is conserved exactly by construction\footnote{This statement assumes that forces and time integration have a zero/negligible error. Obviously, time integration or
approximate forces, e.g. from a gravity solver, can in practice introduce violations of {\em exact} conservation.} as well. Keep in mind that this exact conservation even holds at finite resolution, which is not the case for Eulerian
methods. Moving-mesh methods, see e.g. \cite{springel10b,duffell11}, usually show good overall conservation,
but angular momentum  is not conserved exactly.\\
One can, however, also achieve exact conservation in a more elegant way: by starting from a Lagrangian
and by applying a variational principle \citep{gingold82,speith98,monaghan01,springel02,rosswog09b,price12a} one finds a fully conservative discrete form of the
equations without ambiguities.\\
As before, we will start here with the simplest case of ideal, Newtonian hydrodynamics,
but one can adhere to this strategy also for cases with additional physics, e.g. gravity 
\citep{price07a}. One can start from a discretized Lagrangian of an ideal fluid
\be
L= \sum_b m_b \left( \frac{v_b^2}{2} - u_b \right)
\label{eq:Newtonian_Lagrangian}
\ee
and apply the Euler-Lagrange equations
\be
\frac{d}{dt} \frac{\p L}{\p \vec{v}_a} - \frac{\p L}{\p \vec{r}_a}= 0
\ee
to find the momentum evolution equation of each particle, where $\p L/\p \vec{v}_a$ is 
the canonical momentum, which for Eq.~(\ref{eq:Newtonian_Lagrangian})  simply becomes $m_a \vec{v}_a$. We can use again fixed particle masses and the adiabatic first law of thermodynamics,
$du= -P d(1/\rho)$, which yields
\be
\left(\frac{\p u}{\p \rho}\right)_s= \frac{P}{\rho^2},
\ee 
where $s$ denotes the specific entropy. This yields
\bea
\hspace*{-0.5cm}\frac{d \vec{v}_a}{dt}&=& \frac{1}{m_a}  \frac{\p L}{\p \vec{r}_a}=  - \frac{1}{m_a} \sum_b m_b \frac{\p u_b}{\p \rho_b} \frac{\p \rho_b}{\p \vec{r}_a} = - \frac{1}{m_a}\sum_b m_b \frac{P_b}{\rho_b^2} \left[\sum_k m_k \nabla_a W_{bk}(h_b)\right] \nonumber  \\
&=& - \sum_b m_b \left( \frac{P_a}{\rho_a^2} \nabla_a W_{ab}(h_a) + \frac{P_b}{\rho_b^2} \nabla_a W_{ab}(h_b) \right),
\label{eq:dvdt_NSPH}
\eea
where we have used $\nabla_a W_{bk}= \nabla_b W_{kb} (\delta_{ba} - \delta_{ka})$, see Sec. 2.3 in \cite{rosswog09b}.
For the energy equation, a straight forward insertion of Eq.~(\ref{eq:drho_dt})
into Eq.~(\ref{eq:dudt}) (without the heating/cooling term) yields
\be
\frac{du_a}{dt}= \frac{P_a}{\rho_a^2} \sum_b m_b (\vec{v}_b -\vec{v}_a) \nabla_a W_{ab}(h_a), 
\label{eq:dudt_NSPH}
\ee
and with Eqs.~(\ref{eq:rho_sum}), (\ref{eq:dvdt_NSPH}) and (\ref{eq:dudt_NSPH}) we now
have a complete set of discrete hydrodynamics equations. For practical simulations
one would still need to choose an equation of state so that one can calculate pressures and one has to add dissipative
mechanisms, e.g. a Riemann solver or artificial viscosity, so that shocks can be 
handled robustly.

\subsubsection{General relativistic SPH}
Special- and general-relativistic versions of SPH can be derived  similarly to
the Newtonian case \cite{monaghan01,rosswog09b,rosswog10b,rosswog10a}. We use  geometric units with $G=c=1$ and adopt
(-,+,+,+) for the metric signature of the metric. We reserve greek letters for space-time indices from 0...3
with 0 being the temporal component and we use $i$ and $j$ for spatial components and SPH
particles are labeled by $a,b$ and $k$. Contravariant spatial indices of a vector quantity $\vec{V}$ at particle $a$
are denoted as $V^i_a$, while covariant ones will be written as $(V_i)_a$. The line element and
proper time are given by $ds^2= g_{\mu \nu} \; dx^\mu \; dx^\nu$ and $d\tau^2= - ds^2$ and the proper 
time is related to the coordinate time $t$ by 
\be
\Theta d\tau = dt,
\label{eq:theta_t}
\ee
where we have introduced a generalization of the Lorentz-factor 
\be
\Theta\equiv \frac{1}{\sqrt{-g_{\mu\nu} v^\mu v^\nu}}.
\label{eq:theta}
\ee
The coordinate velocity components are given by
\be
v^\mu= \frac{dx^\mu}{dt}= \frac{dx^\mu}{d\tau} \frac{d\tau}{dt}= \frac{U^\mu}{\Theta}= \frac{U^\mu}{U^0},
\label{eq:v_mu}
\ee
where we have used Eq.~(\ref{eq:theta_t}) and $U^\mu$ is the four-velocity which is normalized,
\be
U^\mu U_\mu= -1. \label{eq:U_norm}
\ee
The Lagrangian of a relativistic fluid  is given by \cite{fock64}
\be
L= - \int T^{\mu \nu} U_\mu U_\nu \sqrt{-g} dV,
\ee
where $g= {\rm det}(g_{\mu \nu})$ and $T^{\mu\nu}$ denotes the energy-momentum tensor which, for an ideal fluid, reads
\be
T^{\mu \nu}= (\tilde{\rho}+P)U^\mu U^\nu + P g^{\mu \nu}.
\ee
Here $\tilde{\rho}$ is the energy density measured in the local rest frame of the fluid and 
$P$ is its pressure. For clarity, we will write
out explicit factors of $c$ in the following lines. The energy density possesses a contribution
from  the rest mass and one from the thermal energy:
\be
\tilde{\rho}= \rho + u \rho/c^2= n m_0 c^2 (1+u/c^2).
\ee
Here $n$ is the baryon number density in the local fluid rest frame, $m_0$ is the baryon mass and $u= u(n,s)$ the specific energy, with $s$ being the specific entropy.\footnote{Here we simply use $m_0= m_u$ for the average baryon mass, where $m_u$ is the atomic mass unit. In reality, $m_0$ depends on the exact composition, but
even in extreme cases the deviations from $m_u$ are only a small fraction of a percent. See Sec.~2.1 in \cite{diener22a} for a more detailed discussion.}
In \SPHI we 
measure all energies in units of $m_0 c^2$.  With this convention (and now using again $c=1$) the energy 
momentum tensor reads
\be
T^{\mu \nu}= \left\{ n (1+u) + P \right\} U^\mu U^\nu + P g^{\mu \nu}.\label{eq:EM_tensor}
\ee
To perform practical simulations, we give up general covariance and choose
a particular frame ('computing frame') in which the simulations are performed. Like in special-relativistic hydrodynamics,
one must clearly distinguish between quantities that are measured in the computing frame
and those measured in the local rest frame of the fluid.
With the normalization of the four-velocity, Eq.~(\ref{eq:U_norm}), the Lagrangian 
can be written as
\be
L= - \int n(1+u)\sqrt{-g} dV\label{eq:Lagrangian_cont}.
\ee
Local baryon number conservation, $(U^\mu n);_\mu= 0$,
can be expressed as 
\be 
\frac{1}{\sqrt{-g}}\p_\mu (\sqrt{-g} U^\mu n)= 0,
\ee
where we have used an identity for the covariant derivative \cite{schutz89}.
More explicitly, we can write 
\be 
\p_t (N) + \p_i(N v^i)= 0, \label{eq:continuity_N}
\ee
where we have made use of Eq.~(\ref{eq:v_mu}) and have introduced the computing frame 
baryon number density
\be
N= \sqrt{-g} \Theta n.\label{eq:N_n}
\ee
In the special relativistic limit $g_{\mu \nu} \rightarrow \eta_{\mu\nu}$ (Minkowski),
$\Theta$ reduces to the special relativistic Lorentz factor $\gamma$ and $\sqrt{-g}$ becomes unity. In this limit, the computing frame density $N_{\rm sr}$ is equal to the local rest frame density $n$ multiplied by the Lorentz factor $\gamma$ between the fluid rest frame and the computing frame. This, of course, makes perfect sense: think of a box in the rest frame of the fluid that contains a certain number of baryons. If this box moves with respect to the computing frame along a coordinate axis, the corresponding box edge appears Lorentz contracted in the computing frame, so that an observer in the computing frame would conclude that the density is
\be
N_{\rm sr}= \gamma n.
\ee
\\
The total conserved baryon number, $\mathcal{N}$, can then be expressed as a sum over fluid parcels
with volume $\Delta V_b$ located at $\vec{r}_b$, where each parcel carries a baryon number\footnote{Be careful not to confuse the baryon number $\nu_b$ with the velocity component of a particle $v^i_b$.} $\nu_b$ 
\be 
\mathcal{N}= \int N dV \simeq \sum_b N_b \Delta V_b = \sum_b \nu_b,\label{eq:parcel_volumes}
\ee
therefore the particle volume in the computing frame reads $\Delta V_b= \nu_b/N_b$.
Eq.~(\ref{eq:continuity_N}) looks like the Newtonian continuity equation
and we will use it for the SPH discretization process. Similar to Eq.~(\ref{eq:kernel_smoothing}), a 
quantity $f$ can now be approximated by 
\be
\langle f \rangle(\vec{r}) \simeq \sum_b \frac{\nu_b}{N_b} \; f_b \; W(\vec{r}-\vec{r}_b,h)\label{eq:SPH_discretization},
\ee
where  the particle's baryon number $\nu_b$ replaces the mass and the Newtonian mass density $\rho$ is replaced by the computing frame baryon number density $N$. The latter can be calculated,
very similarly to the Newtonian case, as
\be
N_a= \sum_b \nu_b W_{ab}(h_a).
\ee
If each particle keeps its baryon number constant, we have
exact baryon number conservation. The fluid Lagrangian, Eq.~(\ref{eq:Lagrangian_cont}), then becomes
\be
L= - \int \frac{1+u}{\Theta} \; N \; dV \approx  - \sum_b  \left( \frac{1+u}{\Theta} \right)_b N_b \Delta V_b = - \sum_b \nu_b \left( \frac{1+u}{\Theta} \right)_b,
\ee
where in the last step we have made use of the volume element $\Delta V_b= \nu_b/N_b$, as suggested
by Eq.~(\ref{eq:parcel_volumes}).\\
Again, as in the Newtonian case, one can use the Euler-Lagrange equations
\be
\frac{d}{dt} \frac{\p L}{\p v^i_a} - \frac{\p L}{\p x^i_a}= 0, \label{eq:EL}
\ee
to derive an evolution equation for the canonical momentum. 
One  can use the Lagrangian to define a canonical momentum 
\be
(p_i)_a\equiv \frac{\p L}{\p v^i_a}= - \frac{\p}{\p v^i_a}\sum_b \nu_b \left( \frac{1+u}{\Theta} \right)_b.
\ee
Note that here $\p u_b/\p v^i_a$ has a non-zero value, since
\be
\frac{\p u_b}{\p v^i_a}= \frac{\p u_b}{\p n_b} \frac{\p n_b}{\p v^i_a} = \frac{P_b}{n_b^2} \frac{\p}{\p v^i_a} \left( \frac{N_b}{\sqrt{-g} \; \Theta_b} \right),
\ee
where we have used the first law of thermodynamics and the
relation Eq.~(\ref{eq:N_n}) and the velocity dependence
comes in via $\Theta$, see Eq.~(\ref{eq:theta}) and one finds \cite{rosswog10a}
\be
\frac{\p}{\p v^i_a} \left( \frac{1}{\Theta_b}\right)= - \Theta_b (g_{i\mu} v^\mu)_a \delta_{ab}.
\ee 
The {\em canonical momentum per baryon} reads
\be
(S_i)_a\equiv \frac{1}{\nu_a} \frac{\p L}{\p v^{i}_{a}}= \Theta_a \left(1+u_a+\frac{P_a}{n_a} \right) (g_{i\mu} v^\mu)_a= \left(1+u_a+\frac{P_a}{n_a} \right) (U_i)_a.
\ee
Similarly, one can use the canonical energy 
\be
E\equiv \sum_a \frac{\p L}{\p v^i_a} v^i_a - L= \sum_a \nu_a \left( v^i_a (S_i)_a + \frac{1+u_a}{\Theta_a}\right)
\ee
to identify the {\em canonical energy per baryon}
\be
e_a \equiv v^i_a (S_i)_a + \frac{1+u_a}{\Theta_a}.
\label{eq:can_energy}
\ee
The evolution equation for the canonical momentum per baryon follows, according to the Euler-Lagrange equations (\ref{eq:EL}), from $\frac{d (S_i)_a}{dt}= \frac{1}{\nu_a} \frac{\p L}{\p x^i_a}$, and after some algebra \cite{rosswog10a} one 
finds\footnote{Note that here we are omitting the "grad-h terms", see \cite{rosswog10a} for the explicit expressions including these small correction terms.}
\bea
\frac{d (S_i)_a}{dt} = - \sum_b  \nu_b 
\left\{
\frac{P_a}{N_a^2} D_i^a  + \frac{P_b}{N_b^2} D_i^b
\right\}
 + \left(\frac{\sqrt{-g}}{2 N} T^{\mu\nu} 
\frac{\p g_{\mu\nu}}{\p x^i}\right)_a,\label{eq:GR_momentum_evolution}
\eea
where we have introduced the abbreviations
\be
D^a_i \equiv   \sqrt{-g_a} \;  \frac{\p W_{ab}(h_a)}{\p x_a^i} \quad {\rm and} \quad 
D^b_i \equiv    \sqrt{-g_b} \; \frac{\p W_{ab}(h_b)}{\p x_a^i}
\label{eq:kernel_grad}.
\ee
The first term in Eq.~(\ref{eq:GR_momentum_evolution}) (including the summation) is due to hydrodynamic accelerations and it is formally very similar to the Newtonian momentum equation (\ref{eq:dvdt_NSPH}), although the involved quantities have a different meaning. The second term represents the metric accelerations from the spacetime curvature and it depends
on the spatial derivatives and the determinant of the metric.\\
The evolution equation of canonical energy per baryon
follows from straight forwardly taking the Lagrangian time derivative of Eq.~(\ref{eq:can_energy}), applying the first law of thermodynamics and after some algebra
\cite{rosswog10a} one obtains
\bea
\frac{de_a}{dt}&=& - \sum_b \nu_b \left\{
\frac{P_a v^i_b}{N_a^2} \; D_i^a  + 
\frac{P_b v^i_a}{N_b^2} \; D_i^b   
\right\}
 - \left( \frac{\sqrt{-g}}{2 N} T^{\mu\nu} \p_t g_{\mu\nu}\right)_a.\label{eq:GR_energy_evolution}
\eea
Similar to the momentum equation, there is a hydrodynamic contribution
(involving the summation over neighbour particles) and a gravitational
contribution that is proportional to the time derivative of the
metric tensor.\\
The attentive reader may wonder how one finds the physical variables (i.e. $v^i, n, u, P$) that are needed in the RHSs of Eqs.~(\ref{eq:GR_momentum_evolution}) and (\ref{eq:GR_energy_evolution}) from the quantities that are actually updated during the integration process, i.e. from $N, S_i$ and $e$. This is actually a not entirely trivial problem (sometimes called "recovery problem", "recovery" or "conservative-to-primitive transformation") that requires a numerical root finding algorithm. The exact algorithm depends on which equations are evolved and on the equation of state that is used. Our strategy in \SPHI is to express $n$ and $u$ in terms of the evolved variables $N,S_i$ and $e$ and the pressure $P$, substitute them into the (polytropic or piecewise polytropic) equation of state and use numerical rootfinding (we use Ridders' method \cite{press92}) to find the pressure value that solves this equation. Once this pressure value is found all physical quantities can be found by a simple backsubstitution. See Sec. 2.2.4 in \cite{rosswog21a} and Appendix A in \cite{rosswog22b} for a detailed description of the polytropic and the piecewise polytropic case, respectively.

\subsubsection{Dissipative terms}
%%%%%%%%%%%%%%%%%%%%%%%%%%%%%%
\begin{figure}[t]
\centerline{
\includegraphics[width=0.7\textwidth]{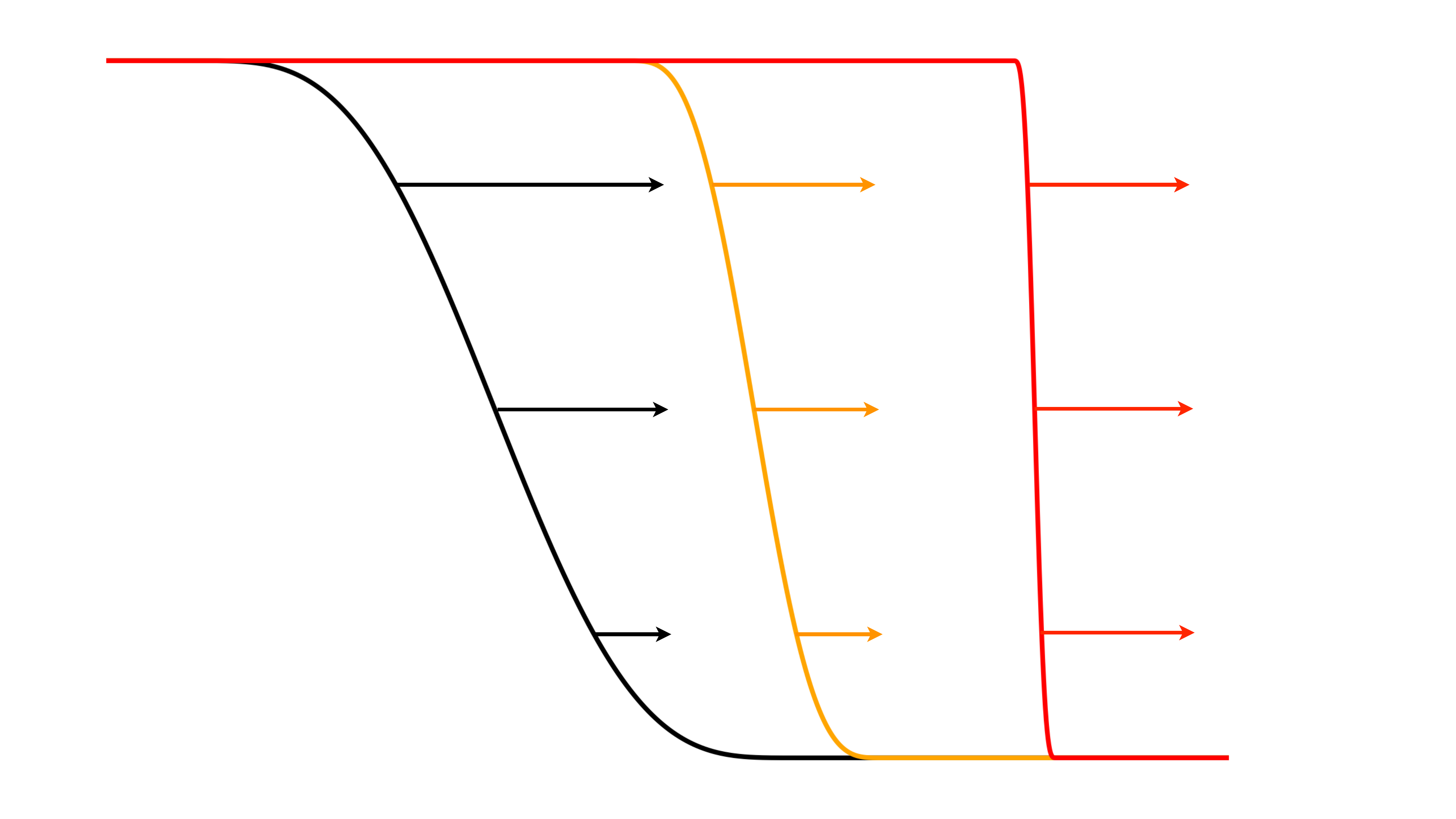}
\includegraphics[width=0.55\textwidth]{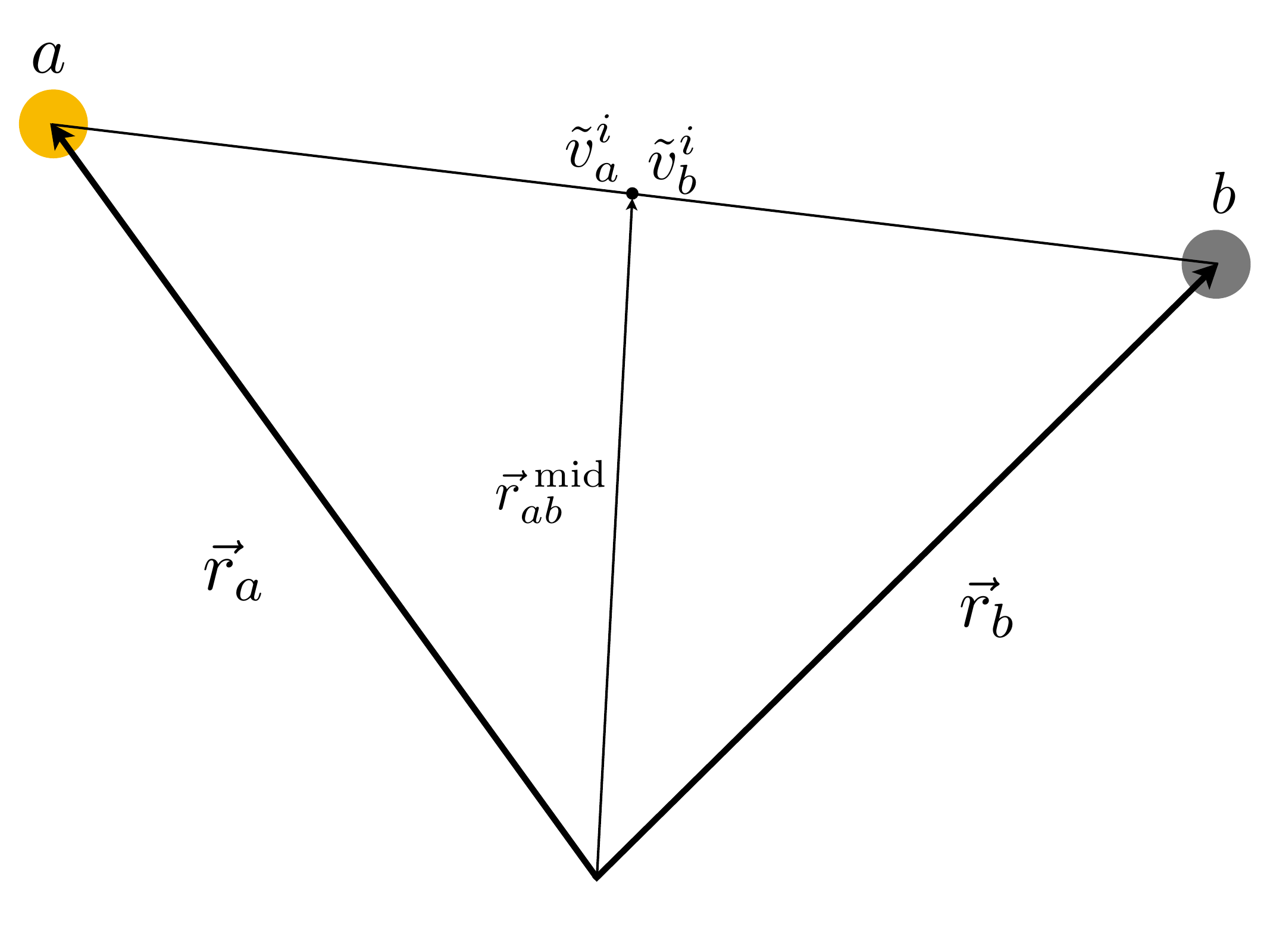}}
\caption{Left: sound wave steepening into a shock. Right: reconstruction of the 
particle velocities to the mid-point between two particles.}
\label{fig:AV}       % Give a unique label
\end{figure}
%%%%%%%%%%%%%%%%%%%%%%%%%%%%%
So far, we have dealt exclusively with strictly non-dissipative hydrodynamics. Most astrophysical gas flows are overall very well described by such ideal fluids, but in realty shocks occur and, in a "microscopic" zoom into the shock front dissipative processes {\em do} occur which lead to an increase in entropy. Numerical methods need to mimick this behaviour, i.e.
in smooth parts of the flow they should (ideally) be dissipationless, but they need to have a built-in mechanism which produces
entropy in a shock front.
 This
can be achieved by either some form of artificial viscosity
or an (approximate or exact) Riemann solver. While both are valid
possibilities \citep{monaghan92,inutsuka02}, we have decided
to use a modern form of artificial viscosity that shares
features with what is usually done in a finite volume scheme \cite{toro99}, in particular, our artificial viscosity treatment
makes use of slope-limited reconstruction.
This new approach to artificial viscosity has been extensively tested in our Newtonian SPH code \Ma \cite{rosswog20a} which served as a technology testbed for \sphi. \\
Artificial viscosity is actually one of the oldest techniques 
of computational physics \citep{vonneumann50}, and while having had
a questionable reputation for some time, it seems that it is becoming 
more popular again, though in more sophisticated forms than originally suggested \citep{cullen10,guermond11,guermond16,rosswog20b,cernetic23,caldana24}. It
is further worth pointing out that modern artificial viscosities share
many similarities with approximate Riemann solvers \citep{monaghan97}.\\
Seemingly harmless sound waves can steepen into shocks during the
hydrodynamic evolution, because the high-density
parts try catch up with the slower low-density parts, see the sketch in the left panel of Fig.~\ref{fig:AV}. Without any dissipation
this process will keep going until the wave has steepened into a mathematical discontinuity
where physical quantities change abruptly. In nature, dissipation
becomes active on some small scale, so that the transition would be very sharp,
usually much sharper than the lengths that can be numerically resolved, but not strictly
discontinuous. The main idea of artificial viscosity is to add an "artificial
pressure" $q$ to the physical pressure $P$ (wherever it occurs)
\be
P \rightarrow P + q
\ee
in order to prevent the wave from becoming infinitely steep and instead 
keep it  at a numerically treatable level. Or, in the words of John von Neumann 
and Robert Richtmyer \citep{vonneumann50}: "{\em Our idea is to introduce (artificial)
dissipative terms into the equations so as to give shocks a thickness
comparable to (but preferentially somewhat larger than) the spacing of
the points of the network. Then the differential equations (more accurately,
the corresponding difference equations) can be used for the entire calculation,
just as if there were no shocks at all}".\\
We will not go into the explicit expressions for the artificial pressure $q$ that we use,
for this we refer to Sec.~2.1.1 in \cite{rosswog22b}, we do, however, want to describe
the basic ideas behind some of the recent artificial viscosity improvements. "Old style" SPH has been criticized
for being overly dissipative. It is worth stating, however, that the SPH method
as derived from the Lagrangian is completely inviscid, and in early SPH implementations
the viscous terms were always switched on.
For pedagogical reasons, we will describe the improvements
at the example of Newtonian SPH where the expressions are simplest, but all
the ideas on how to improve artificial dissipation translate straight forwardly to the GR case 
\cite{rosswog22b}.
In the Newtonian case, the viscous
pressure typically has the form
\be
q_a \propto \alpha_v \; \Delta v_{ab}
\ee
where $\alpha_v$ is a numerical parameter that needs to be of order unity at a shock (but not elsewhere) while $\Delta v_{ab}$ quantifies the "velocity jump" between particle
$a$ and a neighbouring particle $b$ (different expressions for the velocity jump 
are possible).  \\
The major concern is usually to not have dissipation where it is not needed.
There are essentially three  approaches to this. The first one is
to multiply the quantity $\alpha_v$ with a "limiter" that is smart
enough to decide, based on the local fluid properties, whether a particle is
in a shock or not and, in the latter case, the limiter should have 
a very small or zero value to avoid dissipation. Different versions of such limiters have been suggested
\cite{balsara95,cullen10,rosswog15c}. The second approach 
is to make $\alpha_v$ time-dependent \citep{morris97}, so that it
decays exponentially, when it is not needed, to some small, possibly zero value, $\alpha_0$. This approach needs a 
 "trigger" at each particle location that indicates whether more dissipation 
is needed. There are several possibilities for such triggers
\bi
\i The original suggestion of \citet{morris97} was to use $-\nabla \cdot \vec{v}$ as a
trigger so that dissipation switches on during compression. This, however, does not necessarily indicate a shock: gas can also be {\em adiabatically} compressed in which case
no dissipation is wanted.
\i \citet{cullen10} suggested (apart from some other improvements) to construct a 
trigger that is instead based on the time derivative $d (-\nabla \cdot \vec{v})/dt$. This quantity 
indicates a steepening flow convergence which is characteristic
for a particle moving into a shock.
\i \citet{rosswog15b} suggested to use in addition to the {\em shock} trigger of \citet{cullen10}, a {\em noise} trigger that switches on (to a smaller amount), if the flow  becomes numerically noisy, even if there is no shock. Here, the main idea is to measure sign fluctuations of $\nabla \cdot \vec{v}$ in the neighborhood
of a given particle: if there is a clean compression or expansion, all nearby particles
have the same sign of $\nabla \cdot \vec{v}$, but a large number of positive and negative signs,
instead, indicates noise.
\i \citet{rosswog20b} suggested to measure how the entropy of a particle evolves in time.
Since we are modelling an ideal fluid, the entropy should be strictly conserved, this, however,
is not numerically enforced by construction. That means that we can monitor the entropy
conservation as a measure for the numerical quality of the flow and entropy non-conservation (either via shock or numerical noise)  
indicates that dissipation should be applied.
\ei
The third approach to avoid excessive dissipation \cite{rosswog20a}, is actually very similar to 
the reconstruction procedure in Finite Volume schemes \cite{toro99}. Instead of
using the velocity difference between two particles for the velocity jump, 
$\Delta v_{ab}= \vec{v}_a - \vec{v}_b$, one uses instead the differences of the slope-limited
velocities {\em reconstructed to the midpoint} between two particles, $\Delta \tilde{v}_{ab}= \tilde{\vec{v}}_a - \tilde{\vec{v}}_b$, as sketched in
Fig.~\ref{fig:AV}, right panel. The slope-limited,
first order reconstruction from the $a$-side 
reads
\be
\tilde{v}_a^i= v_a^i + \Phi_{ab}(\p_j v_a^i) \; \frac{1}{2} (r_b - r_a)^j 
\ee
and correspondingly for the reconstruction from the $b$-side. For the slope limiter $\Phi_{ab}$ any of the standard
slope limiters can be used in principle. The reconstruction can be extended to higher 
order, for example in the \texttt{MAGMA2} code \cite{rosswog20a} we use a quadratic reconstruction.\\
In \SPHI we have implemented GR-versions of the second and the third of the above ideas. We also evolve our dissipation parameter triggered
by a shock trigger similar to \citet{cullen10}, enhanced
by noise triggers as in \citet{rosswog15b}. In addition,
we perform linear reconstruction to the particle midpoint
in which we use a minmod \cite{roe86} slope limiter. 
Consult Sec.~2.1.1 in \cite{rosswog22b} for the explicit expressions of the artificial viscosity terms used
in \sphi.
\subsection{Spacetime evolution}
In general relativity, gravity is described as curvature of spacetime and the
fundamental object is the spacetime metric, $g_{\mu\nu}$, from which the
spacetime interval between infinitesimally close points can be calculated $ds^2=g_{\mu\nu}dx^{\mu}dx^{\nu}$. 
From the metric tensor we can define the {\em Christoffel symbols}
\be
\Gamma^{\lambda}_{\mu\nu}=\frac{g^{\lambda\sigma}}{2}
\left [\frac{\partial g_{\mu\sigma}}{\partial x^{\nu}}+\frac{\partial g_{\nu\sigma}}{\partial x^{\mu}}-\frac{\partial g_{\mu\nu}}{\partial x^{\sigma}}\right ],
\ee
from which we can calculate the {\em Riemann curvature tensor}
\be
{R^{\lambda}}_{\sigma\mu\nu}=\frac{\partial \Gamma^{\lambda}_{\sigma\nu}}{\partial x^{\mu}}-\frac{\partial \Gamma^{\lambda}_{\sigma\mu}}{\partial x^{\nu}}+\Gamma^{\lambda}_{\rho\mu}\Gamma^{\rho}_{\sigma\nu}-\Gamma^{\lambda}_{\rho\nu}\Gamma^{\rho}_{\sigma\mu}.
\ee
Contracting the first and third index on the Riemann curvature tensor delivers the
{\em Ricci tensor}, $R_{\mu\nu}={R^{\lambda}}_{\mu\lambda\nu}$ and by contracting the two
indices on the Ricci tensor one finds the {\em Ricci scalar} $R=g^{\mu\nu}R_{\mu\nu}$.
These combine to define the {\em Einstein tensor} that is related to the stress-energy
tensor in the full covariant 4-dimensional set of Einstein field equations
\be
G_{\mu\nu}=R_{\mu\nu}-\frac{1}{2}g_{\mu\nu}R=8\pi T_{\mu\nu},
\ee
which, in turn, is the starting point for any spacetime evolution code.\\
%
%%%%%%%%%%%%%%%%%%%%%%%%
\begin{figure}
\centerline{
\includegraphics[width=0.95\textwidth]{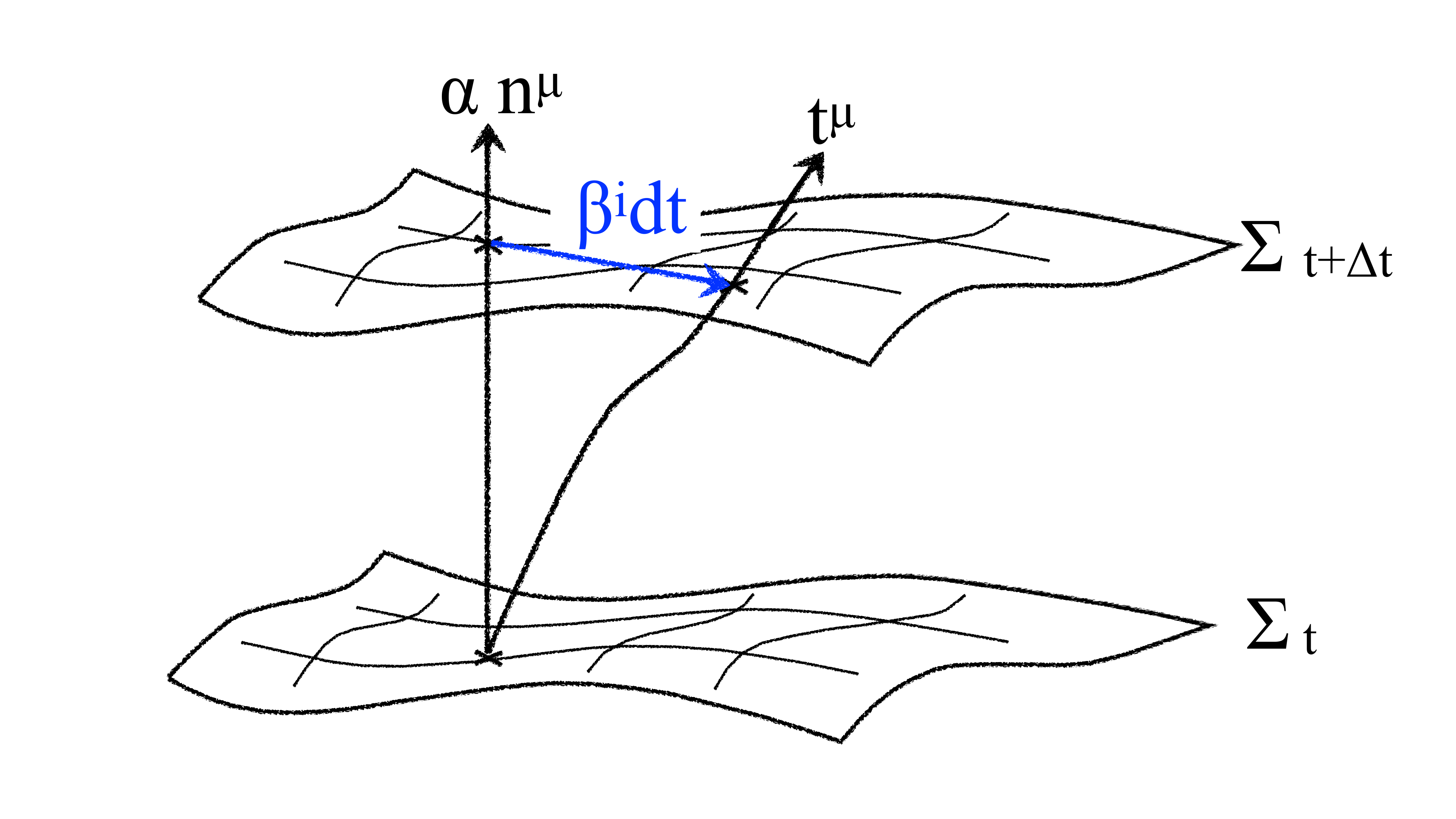}
}
\caption{3+1 foliation of the spacetime. The spacetime is sliced into spacelike hypersurfaces which are integrated forward in time. The lapse function $\alpha$ measures the proper time along the normal $n^{\mu}$, the shift vector $\beta^i$ measures the displacement, on consecutive hypersurfaces, between the observer time lines $t^{\mu}$ and the normal lines $n^{\mu}$.}\label{fig:3+1}
\end{figure}
%%%%%%%%%%%%%%%%%%%%%%%%
These equations are 4-dimensional and it is, therefore, not possible to evolve them
like a normal initial value problem, where one provides 3-dimensional data at a given
time and evolves it forward to the next time. What we can do to achieve this is to split the spacetime into 3+1 dimensions\footnote{This is certainly not the only possible way
to convert the Einstein equations into an initial value problem, but it is a very
common approach and the one we have chosen here.}, that is we write the line element as
\be
ds^2=g_{\mu\nu} dx^{\mu}dx^{\nu}= (-\alpha^2 +\beta_i\beta^i) dt^2+2\beta_i dt dx^i+
     \gamma_{ij}dx^i dx^j,
\ee
where we have introduced the lapse function, $\alpha$, the shift vector, $\beta^{i}$, and the
spatial 3-metric, $\gamma_{ij}$. This foliates spacetime into non-intersecting spacelike hypersurfaces $\Sigma$ labeled by the coordinate
time, $t$, see the sketch in Fig.~\ref{fig:3+1}. Each spatial slice has a future pointing normal vector, $n^{\mu}$, whose 
coordinates are $n^{\mu}=(1/\alpha, -\beta^i/\alpha)$. In this picture the lapse $\alpha$
measures the elapsed proper time between two nearby spatial slices (in the normal direction),
the shift vector, $\beta^i$, is the relative velocity of Eulerian observers and the lines of
constant spatial coordinates and the spatial metric, $\gamma_{ij}$, measure proper distance
within the slices.

The curvature that is intrinsic to the slices is, of course, given by the 3-dimensional Riemann
tensor defined in terms of the 3-metric $\gamma_{ij}$. However, there is also an
extrinsic curvature associated with how the 3-dimensional slices are embedded in the
overall 4-dimensional spacetime. Remarkably, this can be described in terms of another
3-dimensional tensor, $K_{ij}$, that can be defined as the projection of the gradient of
the normal vector onto the spatial slice
\be
K_{\mu\nu}:= -P^{\lambda}_{\mu}n_{\nu;\lambda}, \label{eq:Kij}
\ee
where $P^{\lambda}_{\mu}$ is the projection operator,
\be
P^{\lambda}_{\mu}:=\delta^{\lambda}_{\mu}+n^{\lambda}n_{\mu}.
\ee
Note that even though the
extrinsic curvature is defined as a 4-dimensional object in Eq.~(\ref{eq:Kij}), it is indeed
a 3-dimensional object with $K^{00}=K^{0i}=0$ and can be referred to as $K^{ij}$. 
$K_{00}$ and $K_{0i}$ on the other hand are not zero, but the components of $K_{ij}$ 
can still be obtained by using only the spatial metric, $\gamma_{ij}$, to lower the
indices of $K^{ij}$.

Leaving out the details (see the excellent text books 
\cite{alcubierre08,baumgarte10} for more information) it is then possible to split the 10 original second order
Einstein equations into 12 first order hyperbolic evolution equations
\begin{align}
\dt{\gamma_{ij}} & = -2\alpha K_{ij}+D_i\beta_j+D_j\beta_i, \label{eq:adm1}\\
\dt{K_{ij}} & = -D_i D_j\alpha + \alpha\left [ {}^{(3)}R_{ij}+K K_{ij}-2K_{ik}K^{k}_{j}\right ] \nonumber \\
            & + 4\pi\alpha[\gamma_{ij}(S-\rho)-2S_{ij}]+\beta^k\pdl{K}{ij}{k}+K_{ki}\pdu{\beta}{k}{j}
              + K_{kj}\pdu{\beta}{k}{i}, \label{eq:adm2}
\end{align}
where $D$ is the 3-dimensional covariant derivative related to $\gamma_{ij}$, $\rho=n^{\mu}n^{\nu}T_{\mu\nu}$, 
$S_{\mu\nu}=P^{\lambda}_{\mu}P^{\sigma}_{\nu}T_{\lambda\sigma}$ and $S=S^{\mu}_{\mu}$
and 4 elliptical constraint equations
\begin{align}
    {}^{(3)}R+K^2-K_{ij}K^{ij} & = 16\pi\rho \label{eq:ham} \\
    D_j(K^{ij}-\gamma^{ij}K) & = 8\pi j^i, \label{eq:mom}
\end{align}
where $j^{\mu}=-P^{\mu\lambda}n^{\nu}T_{\lambda\nu}$.
Equations~(\ref{eq:adm1}) and (\ref{eq:adm2}), as written here, are due to 
York~\cite{york79} and are known as the ADM equations after Arnowitt, Deser and 
Misner. Given initial
data for $\gamma_{ij}$ and $K_{ij}$ on a 3-dimensional spatial slice, the ADM
equations can in principle be used to evolve the spacetime forward in time.
Equation~(\ref{eq:ham}) is the {\em Hamiltonian or energy constraint} and 
equation~(\ref{eq:mom}) is the so-called {\em momentum constraint}. These do not involve any time
derivatives but rather provide a set of equations that any physical 
data has to satisfy at any point in time. This is similar, for example, to the case of 
magneto-hydrodynamics, where one has evolution equations
that have to fulfill the $\nabla \cdot \vec{B}=0$-constraint at any point in time.\\
Note that the York form of the ADM equations, also known as standard ADM, differ from 
the original ADM equations~\cite{witten62} by a term proportional to the Hamiltonian 
constraint. Hence they are physically equivalent but not mathematically or numerically
equivalent. This is because the Hamiltonian constraints contains second derivatives 
that changes the principal part and therefore the mathematical properties of the equations.\\
There is a lot of freedom in choosing which of the 12
components of $\gamma_{ij}$ and $K_{ij}$ to provide initial data for and which
of the components to solve for using the constraint equations. Luckily, once
constraint satisfying initial data are set up, evolution of that data using the
ADM evolution equations will guarantee, in the absence of numerical errors, that 
the evolved data remains constraint satisfying. Of course numerical errors are
unavoidable, which means that any numerical evolution will always have some
constraint violations. In that case the constraint violations can be used to
monitor the numerical quality of the simulation.\\
The only problem with the ADM equations is that they do not actually work in practice. They can
be shown (see for example  \cite{alcubierre08}) to be only {\em weakly 
hyperbolic} and hence they are not a well posed formulation of the Einstein equation.
An alternative formulation due to Baumgarte, Shapiro, Shibata and Nakamura, the so-called BSSN formulation \cite{baumgarte10,shibata16},
has proven to be a robust choice for many codes and has
therefore become rather popular. In the BSSN formulation 
a conformal rescaling is introduced
\be
  \tlg_{ij}=e^{-4\phi}\gamma_{ij},
\ee
where $\phi$ is defined in terms of the determinant of the physical three-metric,
$\gamma$, as $\phi = \frac{1}{12}\log\gamma$. With this choice the determinant of the
conformal metric becomes $\tlg=1$. In addition, the extrinsic curvature is separated
into its trace and its tracefree part and conformally rescaled 
\be
  \tlA_{ij}=e^{-4\phi}A_{ij}= e^{-4\phi}(K_{ij}- \frac{1}{3}\gamma_{ij}K).
\ee
The final addition is to also evolve the three quantities
\be
  \tlG^{i}=\tlg^{jk}\tlG^i_{jk}=-\partial_j\tlg^{ij},
\ee
where $\tlG^{i}_{jk}$ is the Christoffel symbols related to the conformal metric.
That is, the BSSN evolution variables are $\phi$, $\tlg$, $K$, $\tlA$ and $\Gamma^{i}$, and result
in the following set of evolution equations
\begin{align}
  \dt{\phi} & = -\frac{1}{6} \left ( \alpha K - \pdu{\beta}{i}{i} \right) + \upwindu{\phi}{}{i}, \label{eq:BSSN1}\\
 \dt{\tlg_{ij}} & = -2\alpha \tlA_{ij} + \tlg_{ik} \pdu{\beta}{k}{j}
                   + \tlg_{jk} \pdu{\beta}{k}{i}
                    -\frac{2}{3} \tlg_{ij} \pdu{\beta}{k}{k}
                         + \upwindl{\tlg}{ij}{k}, \\
  \dt{K} & = -\emfp \left ( \tlg^{ij} \left [ \pdpdu{\alpha}{}{i}{j}
               +2\pdu{\phi}{}{i}\pdu{\alpha}{}{j} \right ]
               - \tlGn^{i}\pdu{\alpha}{}{i} \right ) \nonumber \\
        & + \alpha \left ( \tlA^{i}_{j} \tlA^{j}_{i} +\frac{1}{3} K^2
               \right ) + \upwindu{K}{}{i} + 4 \pi \alpha ( \rho + s ), \label{eq:dtK} \\
  \dt{\tlA_{ij}} & = \emfp \left [ -\pdpdu{\alpha}{}{i}{j} + \tlG^{k}_{ij}
                       \pdu{\alpha}{}{k} + 2 \left ( \pdu{\alpha}{}{i}
                       \pdu{\phi}{}{j}+\pdu{\alpha}{}{j} \pdu{\phi}{}{i}\right ) +\alpha R_{ij} \right ]^{\mathrm{TF}} \nonumber \\
                &      +\alpha ( K \tlA_{ij}- 2 \tlA_{ik} \tlA^{k}_{j} ) + \tlA_{ik} \pdu{\beta}{k}{j}
                       + \tlA_{jk} \pdu{\beta}{k}{i}
                       - \frac{2}{3} \tlA_{ij} \pdu{\beta}{k}{k} \nonumber \\
                 & 
                 +\upwindl{\tlA}{ij}{k} - \emfp \alpha 8 \pi
                       \left (T_{ij}-\frac{1}{3} \gamma_{ij} s\right ), \label{eq:dtA} \\
  \dt{\tlG^{i}} & = -2 \tlA^{ij} \pdu{\alpha}{}{j} + 2 \alpha \left (
                    \tlG^{i}_{jk} \tlA^{jk} - \frac{2}{3} \tlg^{ij}
                    \pdu{K}{}{j} +
                  6 \tlA^{ij} \pdu{\phi}{}{j}\right )
                 \nonumber\\
                  &                     +\tlg^{jk} \pdpdu{\beta}{i}{j}{k} + \frac{1}{3}
                    \tlg^{ij} \pdpdu{\beta}{k}{j}{k} -\tlGn^{j}\pdu{\beta}{i}{j}
                    + \frac{2}{3} \tlGn^{i}\pdu{\beta}{j}{j} \nonumber \\
                    & 
                    + \upwindu{\tlG}{i}{j} -16 \pi \alpha \tlg^{ij} s_j, \label{eq:dtG}
\end{align}
where
\begin{align}
  \rho & =  \frac{1}{\alpha^2} ( T_{00} - 2 \beta^{i} T_{0i} +
             \beta^{i}\beta^{j} T_{ij} ),\label{eq:BSSN_rho} \\
  s & =  \gamma^{ij} T_{ij}, \\
  s_{i} & =  -\frac{1}{\alpha} ( T_{0i} - \beta^{j} T_{ij}),\label{eq:BSSN_Si}
\end{align}
and $\upwindu{}{}{i}$ denote partial derivatives that are "upwinded" based on the
shift vector. This means that the stencil used for
finite differencing is shifted by one in the direction of 
the shift. As an example of this, look at second order
finite differencing, where a derivative in the x-direction 
at grid point $x_i$ would normally be approximated by the 
centered finite difference
\be
\left.\frac{\partial f}{\partial x}\right|_{x=x_i}\approx \frac{-f_{i-1}+f_{i+1}}{2\Delta x} + O\left(\Delta x^2\right).
\ee
If the shift is positive we instead use the "upwinded"
finite difference approximation
\be
\left.\frac{\bar{\partial} f}{\partial x}\right|_{x=x_i}\approx \frac{-3 f_{i}+4 f_{i+1}-f_{i+2}}{2\Delta x}+O\left(\Delta x^2\right), 
\ee
whereas if the shift is negative we use
\be
\left.\frac{\bar{\partial} f}{\partial x}\right|_{x=x_i}\approx \frac{f_{i-2}-4 f_{i-1}+3f_{i}}{2\Delta x}+O\left(\Delta x^2\right).
\ee
The superscript ``TF" in the evolution equation of $\tlA_{ij}$ denotes 
the trace-free part of the bracketed term. Note that there is a
slight subtlety to the treatment of $\tlG^{i}$. We introduce
\be
\tlGn^i=\tlg^{jk}\tlG^i_{jk},
\ee
i.e.\ a numerical recalculation of
the contracted conformal Christoffel symbols from the current
conformal metric. This is used instead of $\tlG^{i}$ whenever
derivatives of $\tlG^{i}$ are not needed. In all other places,
i.e.\ when finite differences are needed, the evolved variables, $\tlG^{i}$, are used directly.
This helps with numerical stability and makes the constraint
$\tlG^{i}=-\partial_j\tlg^{ij}$ better behaved. 
Finally $R_{ij} = \tlR_{ij} + R^{\phi}_{ij}$, where
\begin{align}
  \tlG_{ijk} = & \frac{1}{2}\left ( \pdl{\tlg}{ij}{k} + \pdl{\tlg}{ik}{j}
               - \pdl{\tlg}{jk}{i} \right ), \\
  \tlGmixed{ij}{k} = & \tlg^{kl} \tlG_{ijl}, \\
  \tlG^{i}_{jk} = & \tlg^{il}\tlG_{ljk}, \\
  \tlGn^{i} = & \tlg^{jk} \tlG^{i}_{jk}, \\
  \tlR_{ij}  = &  -\frac{1}{2} \tlg^{kl} \pdpdl{\tlg}{ij}{k}{l}
                  +\tlg_{k(i} \pdu{\tlG}{k}{j)}
                  +\tlGn^{k} \tlG_{(ij)k} %\nonumber \\
            %&   & 
            +\tlG^{k}_{il} \tlGmixed{jk}{l} \nonumber \\
                &  +\tlG^{k}_{jl} \tlGmixed{ik}{l}
                  +\tlG^{k}_{il} \tlGmixed{kj}{l}, \\
  R^{\phi}_{ij} = &  -2\left (\pdpdu{\phi}{}{i}{j}
                 -\tlG^{k}_{ij}\pdu{\phi}{}{k}\right )
                 -2\tlg_{ij} \tlg^{kl} %\nonumber\\
            %& &     
            \left ( \pdpdu{\phi}{}{k}{l}
                 -\tlG^{m}_{kl}\pdu{\phi}{}{m}\right ) \nonumber \\
            &  + 4\pdu{\phi}{}{i}\pdu{\phi}{}{j}
             - 4\tlg_{ij}\tlg^{kl}\pdu{\phi}{}{k}\pdu{\phi}{}{l},
\end{align}
where the standard notation for symmetrization, $A_{(ij)}=(A_{ij}+A_{ji})/2$, has been used.
The additional constraints associated with the conformal metric, $1-\tlg=0$, and
conformal traceless part of the extrinsic curvature, $\tlA=\tlg^{ij}\tlA_{ij}=0$,
are enforced at every substep of the time integrator.\\
Note that a slight variation of BSSN, where $W=\gamma^{-1/6}$ is used as an
evolution variable instead of $\phi$ is implemented in the \SPHI code as well. This
variant seems to be better behaved near the puncture of a black hole, but is
not expected to have a significant effect for binary neutron star mergers.\\
The BSSN evolution equations above do not give a prescription for how to choose the
gauge variables, $\alpha$ and $\beta^{i}$. In order to complete the system, we
add the simplest form of the so-called {\em moving puncture gauges}, i.e.
\be
  \partial_t \alpha = -2 \alpha K
\ee
and
\be
  \partial_t \beta^i = \frac{3}{4}\left (\tlG^{i}-\eta \beta^i\right ).
\ee

\subsection{Particle-mesh interaction}\label{sec:particle_mesh}
Since we are solving the hydrodynamic equations by means of Lagrangian
particles, but evolve the spacetime on a mesh, the particles and the mesh need to continuously exchange information. As one sees from Eqs.~(\ref{eq:GR_momentum_evolution}) and (\ref{eq:GR_energy_evolution}),
the particle evolution equations need the metric and its derivatives,
which are known at grid points, while the spacetime/BSSN-evolution equations, see
Eqs~(\ref{eq:dtK}), (\ref{eq:dtA} and (\ref{eq:dtG}), need the matter's energy momentum tensor, Eq.~(\ref{eq:EM_tensor}),  which is known at the particle positions. In 
other words, the metric and its derivatives need to be interpolated --at every Runge-Kutta substep-- to the particle positions ("mesh-to-particle"-, or M2P-step) and the energy momentum tensor needs to
be mapped to grid points based on the values at the surrounding particles ("particle-to-mesh"-, or P2M-step).\\
The simpler of the two is the M2P-step. In \SPHI it is performed via a 5$^{\rm th}$ order Hermite interpolation, that we have developed in \citep{rosswog21a} 
extending the work of \cite{timmes00a}. Contrary to a standard Lagrange polynomial interpolation, the Hermite 
interpolation guarantees that the metric remains twice differentiable as particles pass from one grid cell to another
and therefore this sophisticated interpolation avoids the introduction of additional noise. 
The details of our approach are explained in Section 2.4 of 
\cite{rosswog21a} to which we refer the interested reader.\\
%%%%%%%%%%%%%%%%%%%%%%%%%%%%%%
\begin{figure}[t]
\centerline{\includegraphics[width=0.99\textwidth]{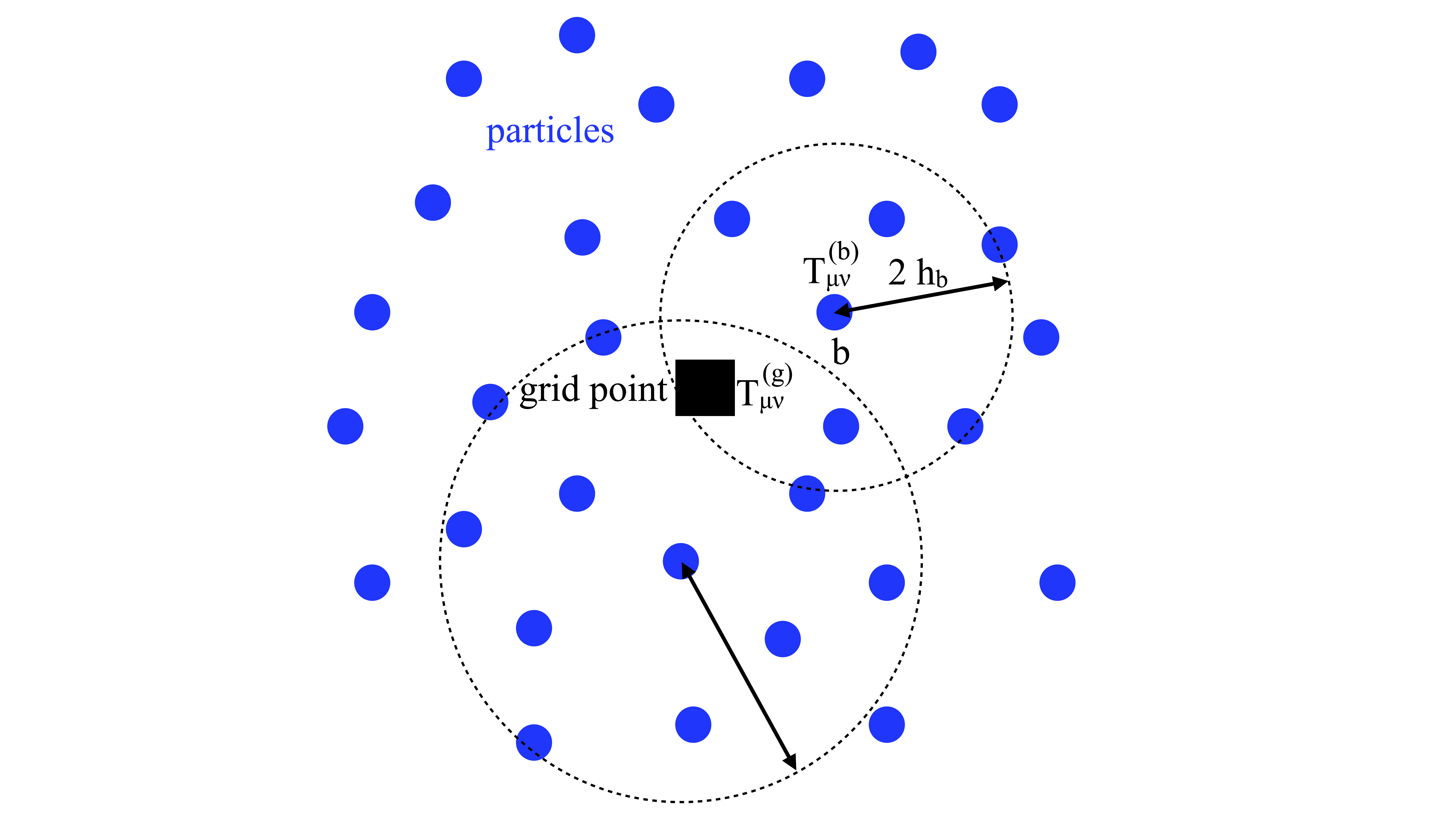}}
\caption{Particle-to-mesh step: the energy momentum tensor $T_{\mu\nu}$ is naturally known at the particle positions (blue circles), since it is needed as a source of the spacetime evolution equations, the particle values need to be "transferred" to the grid points (black square). This is achieved by an LRE-MOOD approach, see the main text for an explanation.}
\label{fig:P2M}       % Give a unique label
\end{figure}
%%%%%%%%%%%%%%%%%%%%%%%%%%%%%
The P2M-step is more complicated since the particle distribution is continuously changing and the particles are in particular not arranged on a regular lattice. Our original approach \cite{rosswog21a,diener22a,rosswog22b} was to use sophisticated kernels that we borrowed from vortex methods \cite{cottet00}, but we have recently \cite{rosswog23a} further refined our P2M-step.
Our new approach is based on a "local regression estimate" (LRE), where for a given polynomial order, we determine the $T_{\mu \nu}$ value on the grid points by the requirement that it is the best fit (according to a suitable error measure) 
to the surrounding particle values, see Fig.~\ref{fig:P2M}. The remaining 
question is then: which polynomial order should be used? Also here  
we follow an optimization approach: we perform trial LRE-mappings for
all orders up to a maximal order (= 4), and out
of those trial mappings we select the one that agrees best with the surrounding particles. This selection process for the polynomial order
is called "Multidimensional Optimal Order Detection" (MOOD), and we therefore
call our P2M-approach  "LRE-MOOD".\\
\subsubsection{LRE-mapping}
The basic idea is to assume that the particle values are given by some 
unknown function $f$, which can be expanded in a Taylor series around a
grid point located at \rg 
\be
f(\vec{r})= f^G + (\p_i f)^G \; (\vec{r} - \vec{r}^G)^i + \frac{1}{2}(\p_{ij} f)^G \; (\vec{r} - \vec{r}^G)^i (\vec{r} - \vec{r}^G)^j + {\rm h.o.t.}. 
\ee
The terms like $(\vec{r} - \vec{r}^G)^i$ can be interpreted as a
polynomial basis $P_l$, shifted to the grid point \rg, with coefficients
$\alpha_l$ that contain the derivatives of $f$.  The 
approximation, "optimized at the grid point G", can then
be written as
\be
\tilde{f}^G(\vec{r})= \sum_{l=0}^{l_{\rm max}} \alpha^G_l P^G_l(\vec{r}).
\ee
Here $l_{\rm max}$ is the number of degrees of freedom and for $d$ spatial dimensions and a maximum polynomial order $m$, it is given by
\be
l_{\rm max}= \frac{(d+m)!}{d! m!}.
\ee
That is we have 1, 4, 10, 20 and 35 degrees of freedom for constant, linear, quadratic, cubic and quartic polynomials. 
We want the $\alpha_l$ to be an optimal fit to the surrounding particles,
therefore we minimize at each grid point the error measure
\be
\epsilon^G\equiv \sum_p \left[f_p - \tilde{f}^G(\vec{r}_p)\right]^2 \; W_{pG},
\ee
where the $f_p$ are the function values at the particle positions and
$W_{pG}$ is a smooth, positive definite function 
(e.g. a typical SPH kernel) that gives particles that are closer to
the grid point a larger weight than to further away particles. The optimal
coefficients, $\alpha^{\rm G}$, are then found from the condition
\be
\left(\frac{\p \epsilon^G}{\p \alpha_i}\right)_G \stackrel{!}{=} 0,
\ee
which yields \cite{rosswog23a}
\be
\alpha_i^G= \left( M_{ik}\right)^{-1} B_k,
\ee
where the "moment matrix" reads
\be
M_{ik}= \sum_p P^G_i(\vec{r}_p) \; P^G_k(\vec{r}_p) \; W_{pG}
\ee
and the vector that contains the information about the function values
reads
\be
B_k= \sum_p f_p \; P_k^G(\vec{r}_p) W_{pG}.
\ee
The moment matrix can be close to singular, therefore we solve for
the $\alpha_i$ by means of a singular value decomposition \cite{press92}.
For more details and some examples of function approximations, we refer to
the appendix of our recent paper \cite{rosswog23a}.

\subsubsection{Multi-dimensional optimal order detection (MOOD)}
While we have just addressed how to determine the best coefficients for 
a given polynomial order, we still need to discuss
how we decide which polynomial order is chosen.
 Here we follow an approach that is sometimes used in computational hydrodynamics \cite{diot13}, where the basic idea is to try a number of possibilities,
dismiss not-admissible solutions (e.g. unphysical ones with $\rho < 0$) and pick the most accurate out of the admissible solutions.  To take this decision, we use the following error measure:
\bea
E^{{\rm G}, m} &\equiv& \sum_p W_{pG} \left[
\sum_{\mu,\nu} \left\{ {\tilde{T}}_{\mu \nu}^{{\rm G}, m} (\vec{r}_p) - T_{\mu \nu,p} \right\}^2 \right]\nonumber \\
&=& 
\sum_p W_{pG} \left[ \sum_{\mu,\nu} \left\{  \left( \vec{\beta}^{G, m}_{\mu,\nu} \cdot P^G(\vec{r}_p) \right)    - T_{\mu \nu,p}   \right\}^2\right],
\label{eq:map_error}
\eea
where $W_{pG}$ is, as before, a smooth positive definite function.
In other words, for each polynomial order $m$, we use the optimal coefficients at the grid point to estimate the function values {\em at each particle position} for those particles that contribute to the grid point and from
the difference between these estimates and the real function values 
at the particle positions we calculate an error measure. The aim is
then to select the polynomial order with the smallest error measure
that is physically admissible. There is, however, one more subtlety....\\

\noindent{\em Identifying the stellar surface}\\
We have tested the just described approach extensively and we noted when simulating neutron stars that in very few cases, 
when particles at the stellar surface contribute to grid cells lying just beneath the surface, the 
lowest error result can be an outlier value compared to the surrounding grid points. Therefore, we 
treat surface grid points differently: here we only apply the lowest polynomial order. This, 
however, requires the identification of which grids points are close to the stellar surface. We do this
by checking which grid cells have many contributions from "surface particles", so we need
to answer the question: {\em How can we identify particles at the stellar surface?}
 \\
The decisive insight to identify the surface
is that at this place "particles are missing on one side" and that this will deteriorate standard SPH-approximations such as 
Eq.~(\ref{eq:better_grad}). If a particle is well engulfed in all directions by other particles, we will find a very good numerical approximation to an analytical result and --vice versa--
a bad approximation indicates that the particle is located near a surface. In practice, we calculate a numerical approximation to $\nabla \cdot \vec{r}=3$
and we use the gradient estimate Eq.~(\ref{eq:better_grad}) for the numerical approximation
\be
\left( \nabla \cdot \vec{r} \right)_a= \sum_b \frac{\nu_b}{N_b} (\vec{r}_b - \vec{r}_a) \cdot \nabla_a W_{ab}(h_a).
\label{eq:divx}
 \ee
From the relative error 
\be
\delta_a \equiv \frac{|(\nabla \cdot \vec{r})_a - 3 |}{3}
\ee
we calculate the average deviation $\langle \delta \rangle_G= (\sum_{b=1}^{n}\delta_b)/n$, where $n$ is the number of particles that contribute to the error measure Equation~(\ref{eq:map_error}).
We use $\langle \delta_G \rangle > 0.05$ as an accurate and robust identification of a
grid point near the fluid surface. For such grid points we only use polynomial order
$m=0$.

\section{Tests and first applications}
\label{sec:applications}
We show here some of the test cases to verify
the correct implementation of the above described methods, more tests can be found
in \cite{rosswog21a,rosswog22b}. We further show some of the first applications
to neutron star mergers.

\subsection{3D shock tube}
%%%%%%%%%%%%%%%%%%%%%%%%
\begin{figure}
\hspace*{-0.5cm}\includegraphics[width=1.05\textwidth]{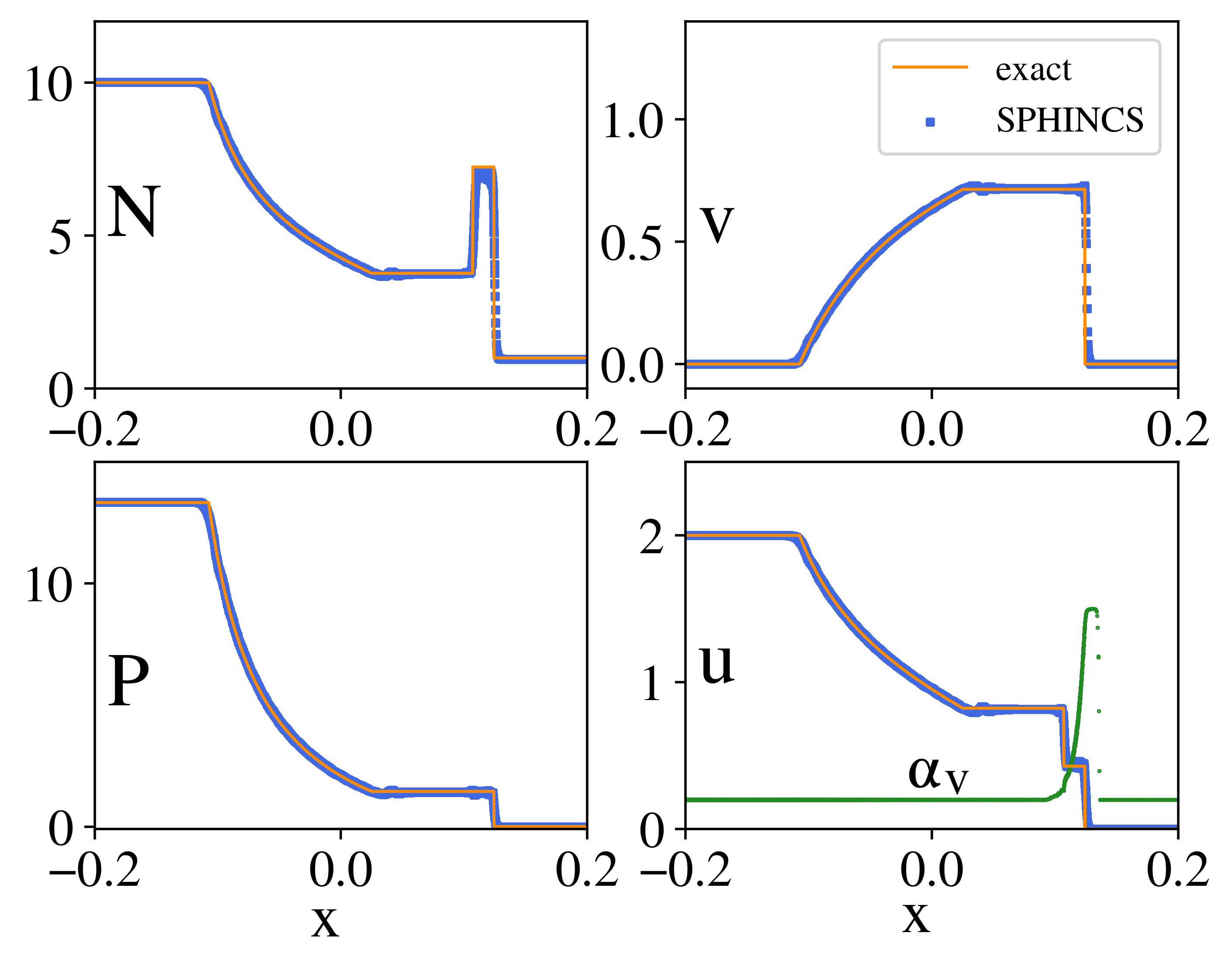}
\caption{3D, relativistic shocktube test. The exact solution is
shown as orange lines, while the numerical results for the computing frame baryon number density $N$, the velocity $v$, the pressure $P$ and the specific internal energy $u$ are shown as blue squares. All particles are shown. We also show in last panel the dissipation parameter $\alpha$ (in green): it "switches on" just directly ahead of the shock, and then very quickly decays to the floor level value.}\label{fig:rel_Sod}
\end{figure}
%%%%%%%%%%%%%%%%%%%%%%%%
With exactly known solutions, shock tubes are excellent tests to verify a correct implementation of the hydrodynamic equations. We perform here a 3D relativistic version 
of ``Sod's shocktube'' \cite{sod78} which has become a widespread benchmark for 
relativistic hydrodynamics codes 
\cite{marti96,chow97,siegler00a,delzanna02,marti03}. The test uses a polytropic 
exponent $\Gamma=5/3$ and as initial conditions 
\be
\left[ N, P \right]=   \left\{
    \begin{array}{ll}
         \left[10,\frac{40}{3}\right], & {\rm for \;}  x<0\\
   	 \left[1,10^{-6}\right]          & {\rm for \;} x \ge 0,
   \end{array}
  \right.
\ee
with velocities initially being zero everywhere. We place  particles with equal baryon numbers on close-packed lattices as described in \cite{rosswog15b},
so that on the left side the particle spacing  is $\Delta x_L= 0.0005$ and we have 12 particles in both $y$- 
and $z$-direction\footnote{The particle spacing on the right side is then given by the
density jump and the requirement that particles have equal baryon numbers.}. This test
is performed with the full 3+1 dimensional code, but using a fixed Minkowski metric.
The result at $t= 0.15$ is shown in Fig.~\ref{fig:rel_Sod} with the \SpB results
marked with blue squares and the exact solution \cite{marti03} with the orange line.
In the last panel we also show the dissipation parameter $\alpha_v$ (in green): it switches on
sharply just {\em ahead} of the shock front and then, once the shock wave has passed, very rapidly decays to
the chosen floor value.
Overall there is very good agreement  with practically no spurious oscillations. 
There is only a small amount of "noise" directly after the shock front.
This is to some extent unavoidable, since here the particles have to re-arrange themselves
from their pre-shock close-packed configuration into their preferred post-shock configuration which goes 
along with some mild sideways motion.

\subsection{Oscillating neutron stars}
In order to test the full code, i.e. the relativistic hydrodynamics,
the spacetime evolution and their mutual coupling via the mappings from the 
grid to the particles and from the particles to the grid, we have performed 
simulations of single oscillating neutron stars. For this problem oscillation 
frequencies are known from independent, linear perturbation approaches \cite{font02,Stergioulas2024} which serve as a measure of the accuracy of our approach.
We solve the Tolman-Oppenheimer-Volkoff (TOV) equations \cite{tolman39,oppenheimer39} for a star
in equilibrium which is then used to set up initial data for the
particles and the spacetime grid. Due to truncation error the star does
not remain in exact equilibrium and different eigenmodes of the star are
excited at eigenfrequencies that depend on the equation of state and the mass
of the star. 
With the choice of $K=100$ and a central density of
$\rho_c=1.28\times 10^{-3}=7.91\times 10^{14}$g/cm$^3$ we get a star with
a gravitational mass of 1.4~\Msun and a baryonic mass of 1.506~\msun.\\
We performed 3 simulations with varying resolution of both the number of
particles and the grid. For the particles we used 500k (low), 1M (medium) and 2M (high)
particles. In all cases we used a grid with outer boundaries (in all directions) at 160 
in code  units, corresponding to 236 km. We used 4 levels of refinement with the star being
completely contained within the finest grid. The resolution on the finest grid was
0.4$\approx$ 590 m (low), 0.317$\approx$ 468 m (medium) and 0.25$\approx$ 369 m 
(high). The smallest smoothing length in each case was about 310 m (low), 255 m 
(medium) and 208 m (high). We evolved the stars up to physical times of almost 30 ms.\\
%%%%%%%%%%%%%%%%%%%%%%%%
\begin{figure}
\hspace*{-1.2cm}\includegraphics[scale=1.8]{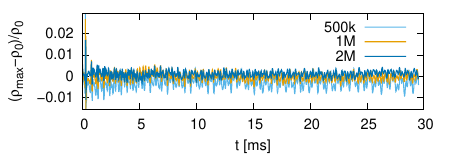}\\
\hspace*{-1.2cm}\includegraphics[scale=1.8]{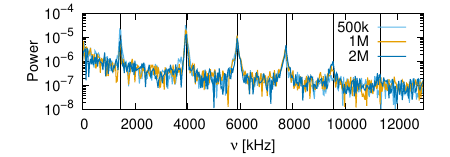}
\caption{Top plot: The relative difference between the maximal density 
$\rho_{\mathrm{max}}$ and the initial maximal density, $\rho_0$, for the 
TOV star as function of time in units of ms. Bottom plot: The power in the 
Fourier spectrum as function of
frequency in units of Hz. In both plots the light blue line is for 500k, the
orange is for 1M and the dark blue is for 2M particles. In the bottom plot, 
the vertical lines indicate (from left to right) the position of the expected frequency for the fundamental mode at 1.442 kHz, the first harmonic at 3.952
kHz, the second harmonic at 5.913 kHz, the third harmonic at 7.771 kHz, the
fourth harmonic at 9.584 kHz and the fifth harmonic at 11.373 kHz. Even
though the fifth harmonic is barely visible, we can still extract the  frequency to better than 1\% accuracy (see Table~\ref{tab:tov}) from
the simulation data.}\label{fig:tov}
\end{figure}
%%%%%%%%%%%%%%%%%%%%%%%%
The results are shown in Fig.~\ref{fig:tov}. In the top plot we show the relative 
difference between the maximal density, $\rho_{\mathrm{max}}$, and the initial maximal 
density, $\rho_0$, as a function of time for the three runs. Here light blue is low, orange
is medium and dark blue is high resolution. As expected, the amplitudes of the oscillations 
decrease with higher resolution ($\approx 0.002$ for 2M particles) and  multiple modes at different 
frequencies are excited. Taking a Fourier transform of
$(\rho_{\mathrm{max}}-\rho_0)/\rho_0$ results in the spectrum shown in the bottom plot of 
Fig.~\ref{fig:tov}. We see one weak peak as well as five strong peaks, all
at frequencies that agree very well with the frequencies of the
fundamental mode and the five first harmonics calculated\footnote{Kindly provided to us by Kostas Kokkotas and Nick Stergioulas.} with the linear  
perturbation code written by Panagiota Kolitsidou~\cite{Stergioulas2024}.
The reference values for the known frequencies are indicated
with black vertical lines in the plot. Quantitative error measures for the
known frequencies are provided in Table~\ref{tab:tov}. Note that even at
the lowest resolution, the frequencies agree to better than 1\% with the 
reference solution and they further improve with resolution.\\
\begin{table}
\caption{Oscillation frequencies: fundamental mode (F), first, second,
third, fourth and fifth harmonic (H1, H2, H3, H4 and H5) frequencies (from
\cite{Stergioulas2024}) are given in Hz in the "Reference" column. For each
resolution we list the extracted frequency of all modes (number before 
$\pm$), as well as an estimate of the accuracy of its extraction (number
after $\pm$). Finally the relative error between the extracted frequency and
the reference frequency is given in the parenthesis. Note that even at the
lowest resolution the agreement of \SPHI results with the reference solution 
is better than 1\%.}
\label{tab:tov}       % Give a unique label
\begin{tabular}{p{1.1cm}p{2.0cm}p{2.5cm}p{2.5cm}p{2.5cm}}
\hline\noalign{\smallskip}
Mode & Reference (Hz) & 500k (Hz) & 1M (Hz) & 2M (Hz) \\
\hline\noalign{\smallskip}
F & 1441.9 & 1435.9 $\pm$ 2.5 (0.4) & 1441.5 $\pm$ 6.1 (0.03) &
1439.0 $\pm$ 4.0 (0.2) \\
H1 & 3952.4 & 3937.0 $\pm$ 3.3 (0.4) & 3942.1 $\pm$ 5.5 (0.3) &
3945.4 $\pm$ 4.2 (0.2) \\
H2 & 5912.5 & 5890.1 $\pm$ 2.7 (0.4) & 5895.7 $\pm$ 5.9 (0.3) &
5901.8 $\pm$ 4.4 (0.2) \\
H3 & 7770.6 & 7713.4 $\pm$ 5.0 (0.7) & 7733.7 $\pm$ 6.1 (0.5) &
7750.4 $\pm$ 4.0 (0.3) \\
H4 & 9583.9 & 9571.5 $\pm$ 32 (0.1) & 9532.4 $\pm$ 52 (0.5) &
9537.8 $\pm$ 21 (0.5) \\
H5 & 11373.1 & 11345.9 $\pm$ 19 (0.2) & 11357.5 $\pm$ 31 (0.1) &
11326.3 $\pm$ 36 (0.4) \\
\end{tabular}
\end{table}
%%%%%%%%%%%%%%%%%
\begin{figure}
\centerline{
\includegraphics[width=0.55\textwidth]{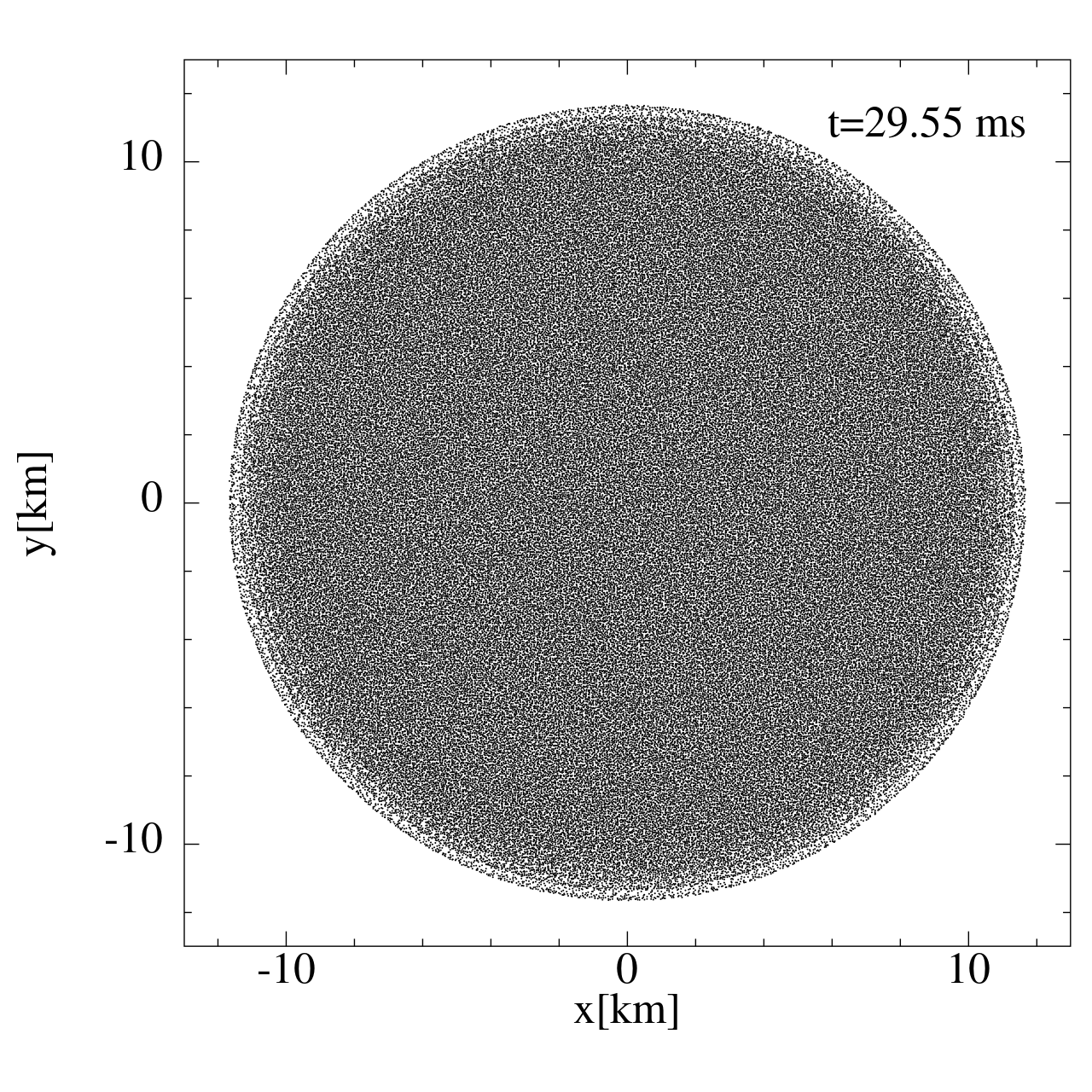}
\includegraphics[width=0.55\textwidth]{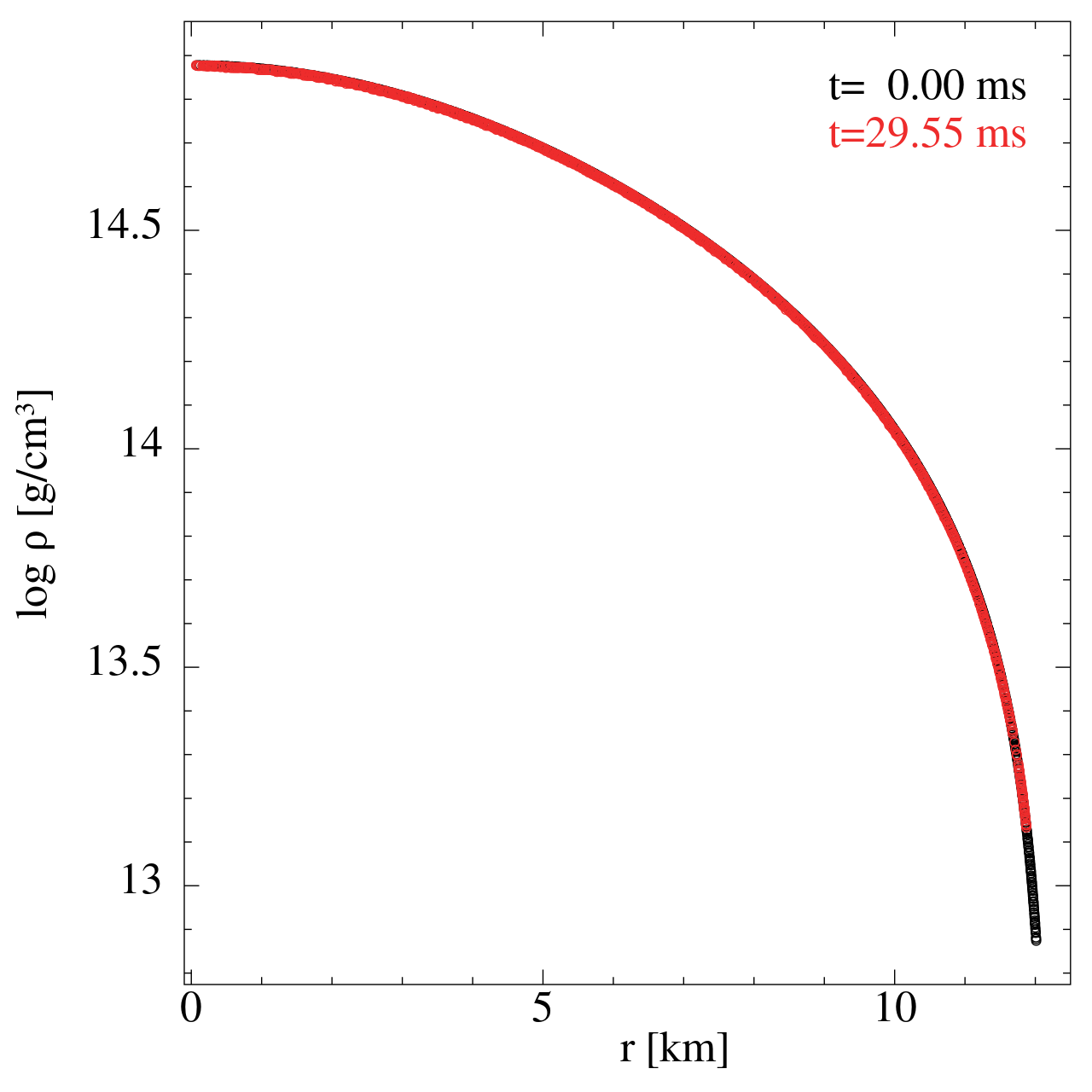}}
\caption{The left plot shows a projection of the particle distribution of a 
2M particle simulation (all particles are shown) at the final time, $t= 29.55$ ms. This time 
corresponds to more then 42 oscillation periods of the fundamental mode. 
Note that the neutron star surface has remained perfectly well behaved 
and the density distribution at late time (red) is nearly identical
to the initial condition (black).
 }\label{fig:tov_2M}
\end{figure}
In Fig.~\ref{fig:tov_2M} we show a projection to the $xy$-plane of the particle distribution at the end
of the simulation (left plot) and a comparison between the initial (black) and final 
(red) density as a function of radius (right plot). The final time, $t=29.55$ ms, 
corresponds to more than 42 oscillation periods at the fundamental mode frequency.  Note
that, despite the long evolution time, the surface of the TOV star is extremely well behaved with no "outlier" particles (the plot
shows all the particles). In contrast, the 
neutron star surface in {\em Eulerian} simulations is a perpetual source of numerical trouble where often (completely spurious) "winds" are driven away from the star.
From the right plot it is clear that the density
distribution has changed very little over the duration of the simulation. The main
difference is that particles near the surface have moved inward by a very small amount compared
to the starting position, but overall the density distribution has remained very well-behaved and is essentially identical to the
initial condition.
\subsection{Neutron star mergers}
One of the key motivations behind the \SPHI development efforts are the multi-messenger signatures
of merging neutron star binaries. We will show here
two examples, one merger where the remnant remains stable until the end of the simulation and another one
where the encounter results in a prompt collapse to a black hole.
%%%%%%%%%%%%%%%%%%%%%%%%
\begin{figure}
\centerline{
\includegraphics[width=1.2\textwidth]{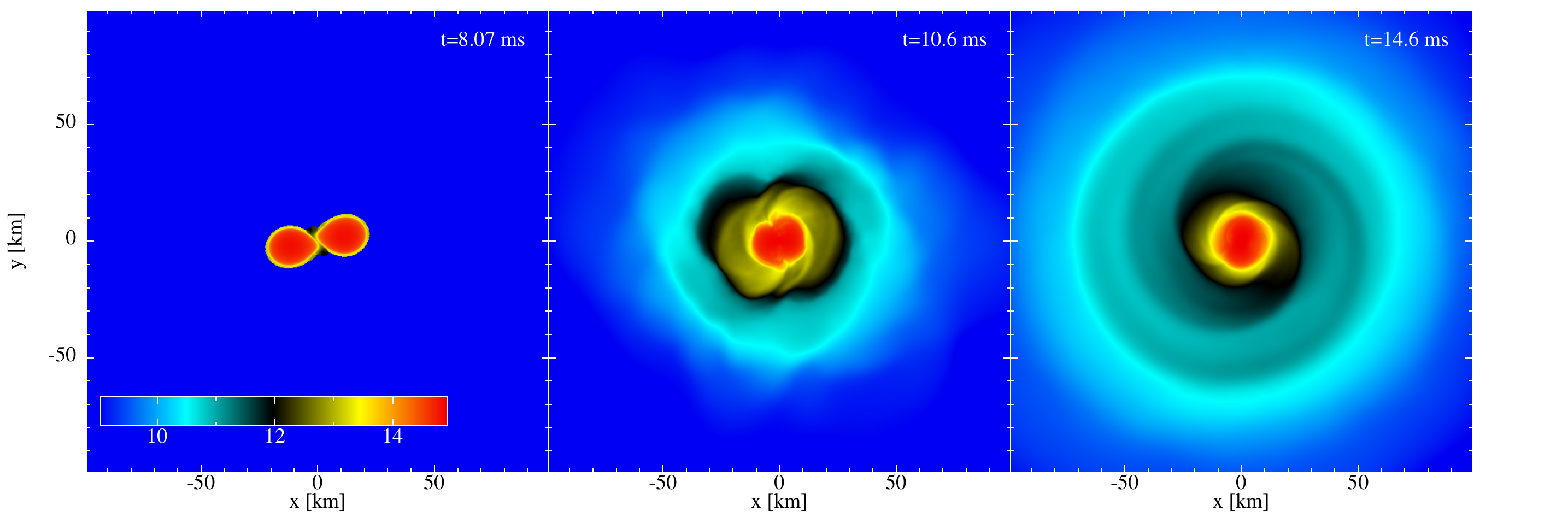}}
\caption{Merger of an irrotational binary system with $2 \times 1.3$ \Msun with the APR3 equation of state. Shown is the density distribution in the orbital plane.}\label{fig:density_APR3}
\end{figure}
%%%%%%%%%%%%%%%%%%%%%%%%
\subsubsection{Merger with stable remnant}
\label{sec:stable}
\begin{figure}

\hspace*{-1.0cm}\includegraphics[scale=1.8]{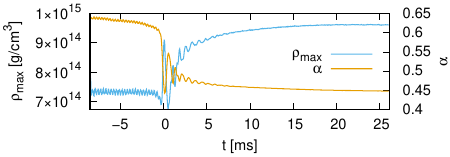}\\
\hspace*{-1.0cm}\includegraphics[scale=1.8]{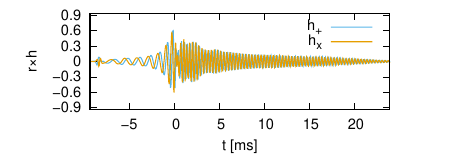}
\caption{Top: evolution of the peak density and minimum lapse value during 
the merger of the 2$\times$1.3 \Msun merger with APR3 equation of state. Bottom: corresponding gravitational wave amplitudes (multiplied with the distance to the source).}\label{fig:rho_alpha}
\end{figure}

As an example of a neutron star merger that results in a surviving remnant, we show a binary of  2$\times$1.3 \Msun  with the piecewise polytropic representation \cite{read09} of the APR3 equation of state \cite{akmal98}, where we add a thermal
pressure/internal energy component with a thermal polytropic exponent $\Gamma_{\rm th}= 1.75$, see Fig.~\ref{fig:density_APR3}.
The spacetime/matter initial conditions for this binary system were calculated using the FUKA code \cite{papenfort21}, subsequently the particles were placed according to the "artificial pressure method" \cite{rosswog20a,rosswog23a} with the code SPHINCS$\_$ID.
We start from an initial separation $a_0$= 45 km and use 2 million SPH particles to model the matter evolution while for
the spacetime evolution we initially employ seven levels of mesh refinement with 193 points in each direction. During the merger and the subsequent evolution, the spacetime
becomes more extreme and \SPHI automatically increases the number of refinement
levels to eight, see Sec.~2.5 in \cite{rosswog23a} for more information on our mesh refinement algorithm.\\
Fig.~\ref{fig:density_APR3} shows the density evolution (in the orbital plane) and in the top of Fig.~\ref{fig:rho_alpha} we show how the maximum density and the minimum value of the lapse function
evolve. As the stars approach each other, the lapse continuously 
decreases, while the peak density stays essentially constant.
The merger results in a strong initial compression (by $\sim$ 25\% in the density), a subsequent "bounce back" during which the peak density drops below the initial single-star density, and then a continuous increase which levels off to 
a constant value at $\sim$ 15 ms after the  merger. The evolution of the lapse function during the merger is anti-correlated with the peak density: during maximum compression the laps drops to a minimum value of 0.42 and reaches after $\sim 15$ ms an asymptotic value of 0.45.\\
In the bottom plot of Fig.~\ref{fig:rho_alpha} we show, in light blue, the plus- and,
in orange, the cross-polarization of the strain times the distance to the source, $r$. The gravitational waves has been extracted from the simulation via the Newman–Penrose 
Weyl scalar $\Psi_4$ at an extraction sphere of 
$R_{\mathrm{extr}}=150\approx 221.5$ km, see the appendix A of \cite{diener22a} for more
details and we have used kuibit~\cite{bozzola21} for their analysis. At the earliest stages ($t <-7$ ms; $t= 0$ marking the peak GW emission) there is a small spurious transient introduced by
the initial conditions, subsequently one recognizes the expected "chirp" signal with a long ring-down phase.

%%%%%%%%%%%%%%%%%%%%%%%%
%%%%%%%%%%%%%%%%%%%%%%%%

\subsubsection{Merger with prompt black hole formation}
As an example of a system that undergoes prompt collapse to a black hole, we show
the merger of a  2$\times$1.5 \Msun binary with the same equation of state, initial
separation, particle number and initial grid setup as in Sec.~\ref{sec:stable}.
As this system is massive enough to undergo prompt collapse to a black hole, more
refinement levels are added automatically as the collapse proceeds until, at the end,
we have 11 levels of refinement. Again see Sec.~2.5 in \cite{rosswog23a} for the 
details.   
\begin{figure}
\hspace*{-1.0cm}\includegraphics[scale=1.8]{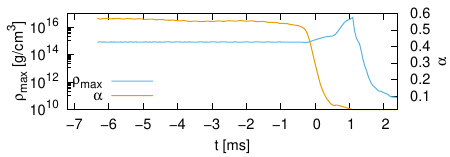}\\
\hspace*{-1.0cm}\includegraphics[scale=1.8]{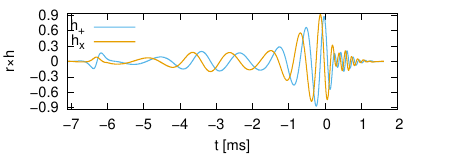}
\caption{Top plot: we show in blue (left axis) the maximal density, 
$\rho_{\mathrm{max}}$, and in orange (right axis) the lapse, $\alpha$, as function
of time  for the system with two 1.5 \Msun neutron stars with the
APR3 equation of state. Bottom plot: we show in blue the plus-polarization, $h_+$,  
and in orange the cross-polarization, $h_{\times}$, of $r$ times the gravitational wave strain.\label{fig:collapse}
}
\end{figure}
In the top plot of Fig.~\ref{fig:collapse} we show the maximal density,
$\rho_{\mathrm{max}}$, (light blue, left axis), and the lapse, $\alpha$, (orange,
right axis) as function of time. As can be seen the lapse starts collapsing right
after the merger while the maximal density starts to increase dramatically. 
The collapsing remnant requires a rapid decrease in the numerical time step as described in Sec.~2.5 in
\cite{rosswog23a}. Once the lapse value of a particle
decreases below 
 $\alpha_{\mathrm{dust}}=0.05$, we convert it to "dust", which means that we ignore the particle's internal energy and pressure contribution. This makes the conservative-to-primitive recovery trivial. Below $\alpha_{\rm cut}= 0.02$ we remove particles
 from the simulations. As can be double-checked by calculating a posteriori the apparent horizon, this procedure is safe and does not lead to notable artifacts.
During the collapse we reach a maximal density of
$\rho_{\mathrm{max}}=5\times 10^{16}$ g/cm$^3$ before the removal of particles kicks in. At the
end of the simulation about $7\times 10^{-4}$ \Msun of material are still
outside the black hole out of which about $2\times 10^{-4}$ \Msun are unbound, and 
some fraction of which has speeds exceeding $0.7 c$. The remaining bound material is expected to fall into the black hole eventually.

In the bottom plot of Fig.~\ref{fig:collapse} we show, in light blue, the plus- and,
in orange, the cross-polarization of $r$ times the strain extracted from the
simulation at an extraction sphere of $R_{\mathrm{extr}}=150\approx 221.5$ km.
Note that, in both the bottom and the top plot, the data is shifted so that the time
of maximal gravitational wave power corresponds to $t=0$. The waveform 
initially ($-7<t<-5.5$ ms) shows some spurious
waves from the initial data, but then shows a clean "chirping" inspiral, the merger ($-0.5<t<0.6$ ms) and, finally, the black hole ringdown ($t>0.6$ ms). 

\begin{figure}
\hspace*{-1cm}\includegraphics[width=1.1\textwidth]{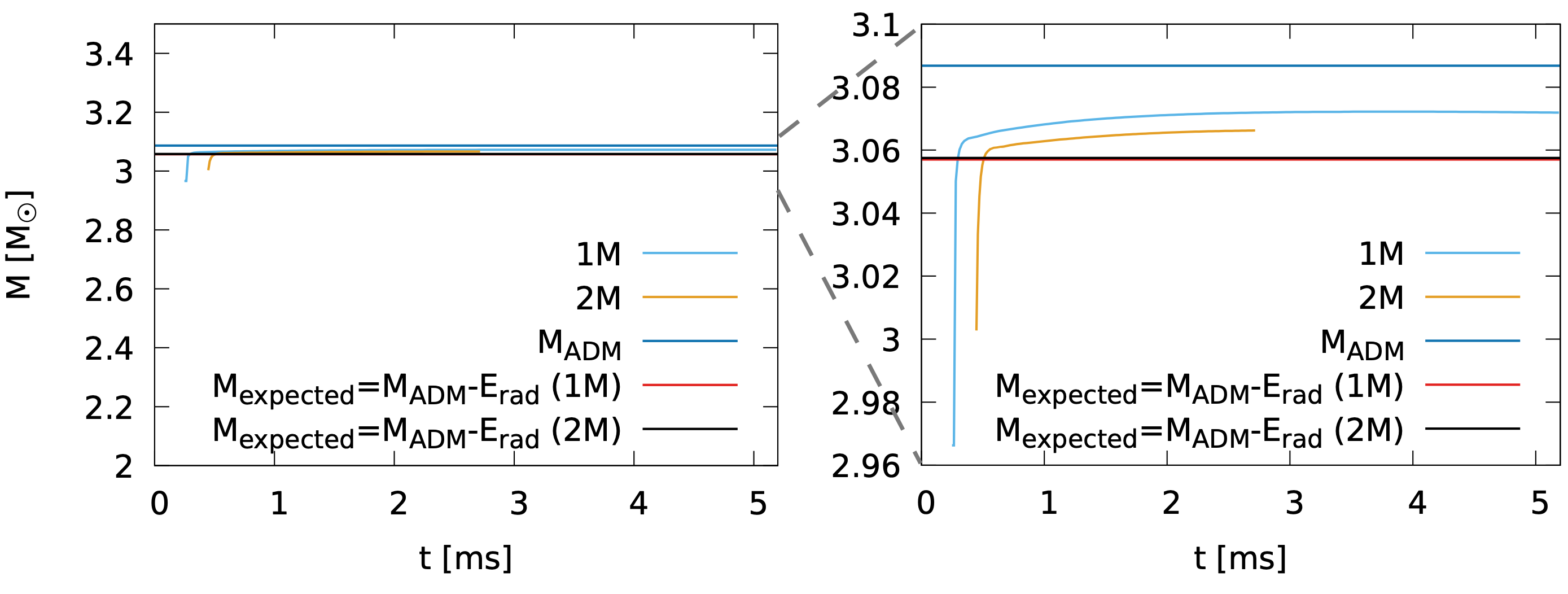}
\caption{The mass of the black hole (including rotational energy) as function of time 
(the time has been shifted so zero corresponds to the time of maximal gravitational 
wave power; left panel is zoom-in of right panel). The orange curve is for the 2M particle runs mentioned earlier, while the 
light blue curve is for a lower resolution run with 1M particles. The dark blue curve 
shows the initial ADM mass of the system, while the black and red curves shows the 
initial ADM mass minus the radiated gravitational wave energy, i.e.\ the expected
final mass of the black hole, as determined by the 2M and 1M runs, respectively.}\label{fig:tot_mass}
\end{figure}
We have written a reader for spacetime data for the Einstein Toolkit~\cite{loeffler12} to analyze the properties of the black hole that forms. We first
used \AHFD~\cite{Thornburg95,Thornburg03} to find the apparent horizon and, once found, \QLM~\cite{Dreyer:2002mx,Schnetter:2006yt} to calculate the spin and
total mass of the black holes. The result is shown in Fig.~\ref{fig:tot_mass}. Here 
we show the 
total mass of the black hole for the 2M particle run in orange, the
result of a lower resolution 1M particle run is shown in light blue. As can be seen
the mass of the black hole grows rapidly at first then slows down and eventually levels
off. The dark blue line shows the ADM mass of the initial data, while the black and red
lines shows the initial ADM mass\footnote{In General Relativity energy and momentum are non-local quantities and difficult to define since the spacetime itself contributes \cite{misner73}. There are various definitions of "mass", the ADM mass applies to asymptotically flat spacetimes and can be found by integrating over the spacetime.} minus the energy radiated away in gravitational waves
as calculated from the 2M (black) and 1M (red) runs. This is the expected final mass,
$M_{\mathrm{final}}$, of the black hole. Both cases produce final black hole masses very close to this value with the 2M run reproducing it to within 0.3 \%. For even better agreement, higher resolution would be needed. In the higher resolution run, the final dimensionless spin parameter
is $a/M\approx 0.75$.

\section{Summary and outlook}
\label{sec:summary}
We have presented here the main methodological elements of the {\em Lagrangian} numerical
relativity code \sphi. This code evolves the spacetime in a very similar fashion to standard
{\em Eulerian} codes, in particular by solving the BSSN equations \cite{alcubierre08,baumgarte10} 
on an adaptive mesh. The major difference to more conventional numerical relativity codes is that the fluid is evolved by means of Lagrangian particles. This has major
advantages for tracking and evolving the matter that is ejected in a neutron star 
merger. This matter, although only a relatively small fraction ($\sim$1\%) of the binary mass, is responsible for all the electromagnetic emission and therefore
of paramount importance for predicting the multi-messenger signatures on compact
binary mergers. This mixed methodology with both particles and adaptive mesh, however, requires an accurate information exchange between the particles and the grid points and this has been
one of the major challenges of the development process. As described in Sec.~\ref{sec:particle_mesh}, we have found a way that produces accurate and
robust results.\\
To demonstrate the proper functioning of the hydrodynamic part of \SPHI we have 
performed 3D shock tube tests which show very good agreement with the exact 
solution and thereby demonstrate the accurate functioning of the hydrodynamic part. Neutron star oscillations where the full spacetime is evolved are 
a good test for scrutinizing both the spacetime evolution and the 
coupling of spacetime and hydrodynamics, since measured
oscillation frequencies can be compared against accurately known values from independent approaches. We have evolved
a neutron star  for $\sim 30$ ms and measured its oscillation frequencies.
We find excellent agreement with the results of \cite{font02}: even at low resolution
all frequencies agree already to better than 1\% and become increasingly more
accurate with higher resolution. The neutron star remains essentially perfectly
in its initial state and its surface remains very well-behaved without any artifacts.
This is a major advantage of \SPHI compared to Eulerian codes.
We have further performed two binary neutron star simulations, one where a stable remnant survives and another one where a black hole forms promptly.\\
%%%%%%%%%%%%%%%%%%%%%%%%
\begin{figure}
\centerline{
\includegraphics[width=0.50\textwidth]{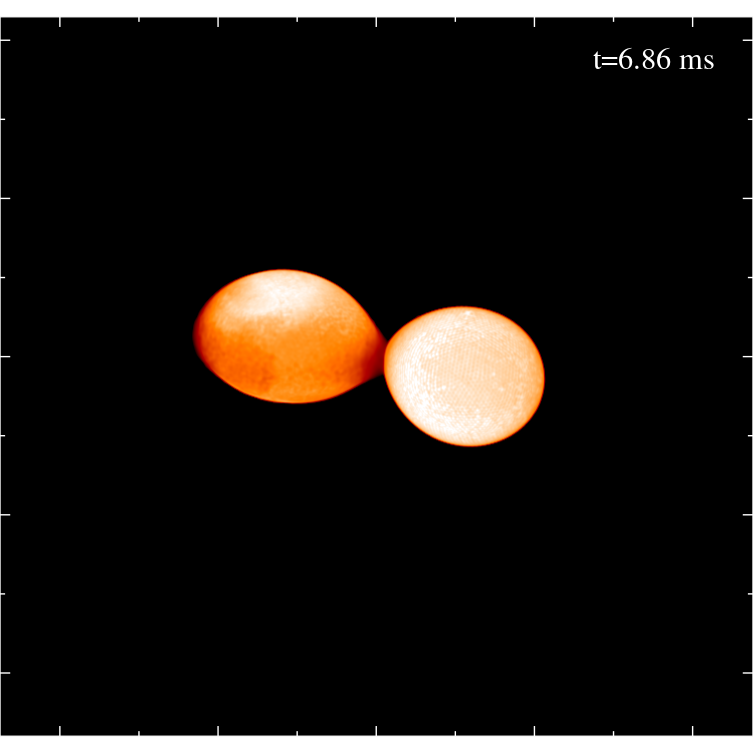}
\includegraphics[width=0.50\textwidth]{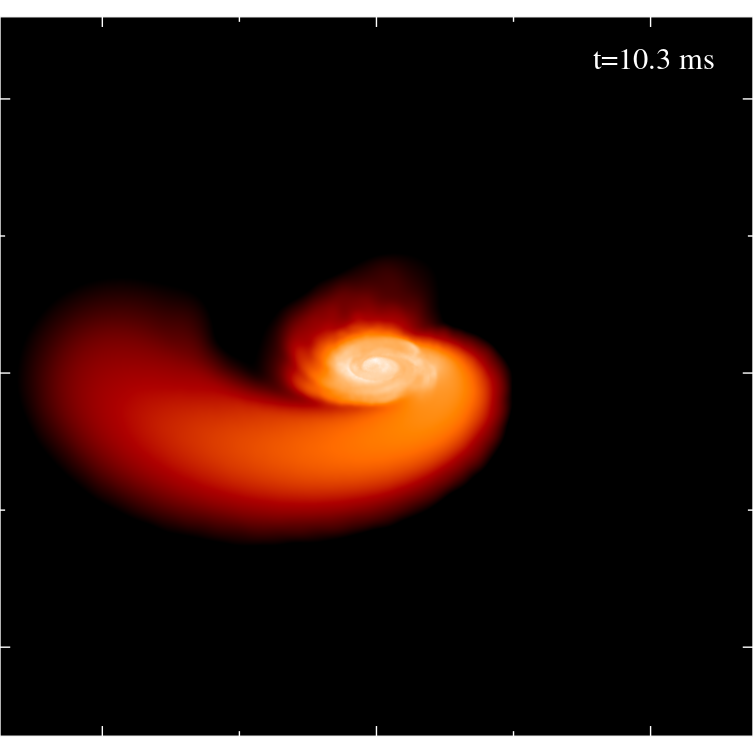}
}
\caption{Volume rendering of an equal mass ($2\times1.3$ \msun) binary merger where one of the stars (the left star in the left panel) is rapidly spinning ($\chi_1=0.5$) while the other is not ($\chi_2=0$). The multi-messenger signatures of such systems are explored in detail in \cite{rosswog24b}.}\label{fig:volren_spin}
\end{figure}
%%%%%%%%%%%%%%%%%%%%%%%%
After an extensive phase of scrutinizing and further improving our methodology,
we have recently begun to simulate astrophysical systems \cite{rosswog24a}. Fig.~\ref{fig:volren_spin}
shows a example of a neutron star binary ($2\times1.3$ \msun, APR3 EOS) where one
star is rapidly spinning with dimensionless spin parameter $\chi_1= 0.5$ while
the other star has no spin ($\chi_2=0$). Such binaries with only one millisecond neutron star are not just a mere
academic possibility, but they likely form in dense stellar environments such
as globular clusters via binary-single exchange encounters \cite{rosswog24b}. The large majority of neutron stars observed in globular clusters are rapidly spinning\footnote{See P. Freire's website https://www3.mpifrbonn.mpg.de/staff/pfreire/NS\_masses.html} and encounters between a binary, consisting of a spun-up neutron star and a low-mass companion, and another neutron star can easily lead to the exchange
of the low mass companion with the neutron star. According to our estimates \cite{rosswog24b}, such binaries may account for approximately $\sim 5$\% of the merging neutron star systems.
Such mergers show a number of distinct signatures in their multi-messenger signals:
they dynamically eject approximately an order of magnitude more matter and therefore produce particularly bright kilonovae. Since the collision itself is less violent they 
produce a smaller amount of fast ejecta ($>0.5c$) and therefore weaker blue/UV kilonova precursors \cite{metzger15a}. The GW-emission is substantially less efficient and therefore one can expect a longer lived central remnant and one may speculate that such mergers may be related to long GRBs that produce kilonova emission such as GRB 211211A \cite{rastinejad22} and GRB 230307A \cite{levan24}.\\
Future development efforts of \SPHI include improvements of its parallelization
and the implementation of further physics ingredients such as, for example, more realistic nuclear matter equations of state and neutrino transport.

\begin{acknowledgement}
It is a great pleasure to acknowledge interesting discussions with Sam Tootle and we are very grateful for his generous help with the FUKA library. We are further grateful to Kostas Kokkotas and Nick Stergioulas for sharing their insights into oscillating stars and we would also like to thank Diego Calderon, Jan-Erik Christian, Pia Jakobus, Laurens Paulsen, Lukas Schnabel and Wasif Shaquil for their careful reading of an earlier version of this text. 
SR has been supported by the Swedish Research Council (VR) under 
grant number 2020-05044, by the research environment grant
``Gravitational Radiation and Electromagnetic Astrophysical
Transients'' (GREAT) funded by the Swedish Research Council (VR) 
under Dnr 2016-06012, by the Knut and Alice Wallenberg Foundation
under grant Dnr. KAW 2019.0112,   by the Deutsche 
Forschungsgemeinschaft (DFG, German Research Foundation) under 
Germany's Excellence Strategy - EXC 2121 ``Quantum Universe'' - 390833306 
and by the European Research Council (ERC) Advanced 
Grant INSPIRATION under the European Union's Horizon 2020 research 
and innovation programme (Grant agreement No. 101053985). \\
Parts of the simulations for this chapter have been performed on the facilities of
 North-German Supercomputing Alliance (HLRN), and at the SUNRISE 
 HPC facility supported by the Technical Division at the Department of 
 Physics, Stockholm University. Special thanks go to Holger Motzkau 
 and Mikica Kocic for their excellent support in upgrading and maintaining 
 SUNRISE.
 Portions of this research were conducted with high performance computing resources provided by Louisiana State University (http://www.hpc.lsu.edu).

\end{acknowledgement}

\bibliographystyle{aipnum4}
\bibliography{astro_SKR,nr_PD}

\end{document}